# The dynamical evolution of dwarf planet (136108) Haumea's collisional family: General properties and implications for the trans-Neptunian belt

**Patryk Sofia Lykawka,[1*] Jonathan Horner,[2] Tadashi Mukai,[3] Akiko M. Nakamura[4]**

[1] Astronomy Group, Faculty of Social and Natural Sciences, Kinki University, Shinkamikosaka 228-3, Higashiosaka-shi, Osaka, 577-0813, Japan
[2] Dept. of Astrophysics, School of Physics, University of New South Wales, 2052, Australia
[3] Kobe University, 1-1 rokkodai-cho, nada-ku, Kobe 657-8501, Japan
[4] Dept. of Earth and Planetary Sciences, Kobe University, 1-1 rokkodai-cho, nada-ku, Kobe 657-8501, Japan

patryksan@gmail.com
http://sites.google.com/site/patryksofialykawka/
*******

[*]E-mail address: patryksan@gmail.com

**ABSTRACT**
Recently, the first collisional family was identified in the trans-Neptunian belt (otherwise known as the Edgeworth-Kuiper belt), providing direct evidence of the importance of collisions between trans-Neptunian objects (TNOs). The family consists of the dwarf planet (136108) Haumea (formerly 2003 EL61), located at a semi-major axis, $a$, of ~ 43AU, and at least ten other ~100km-sized TNOs located in the region $a$ = 42 - 44.5 AU. In this work, we model the long-term orbital evolution (4 Gyr) of an ensemble of fragments (particles) representing hypothetical post-collision distributions at the time of the family's birth based on our limited current understanding of the family's creation and of asteroidal collision physics. We consider three distinct scenarios, in which the kinetic energy of dispersed particles were varied such that their mean ejection velocities ($v_{eje}$) were of order 200 m/s, 300 m/s and 400 m/s, respectively. Each simulation considered resulted in collisional families that reproduced that currently observed, despite the variation in the initial conditions modeled. The results suggest that 60-75% of the fragments created in the collision will remain in the trans-Neptunian belt, even after 4 Gyr of dynamical evolution. The surviving particles were typically concentrated in wide regions of orbital element space centred on the initial impact location, with their orbits spread across a region spanning $\Delta a$ ~ 6-12 AU, $\Delta e$ ~ 0.1-0.15 and $\Delta i$ ~ 7-10°, with the exact range covered being proportional to the $v_{eje}$ used in the model. Most of the survivors populated the so-called Classical and Detached regions of the trans-Neptunian belt, whilst a minor fraction entered the Scattered Disk reservoir (<1%), or were captured in Neptunian mean motion resonances (<10%). In addition, except for those fragments located near strong resonances (such as the 5:3 and 7:4), the great majority displayed negligible long-term orbital variation. This implies that the orbital distribution of the intrinsic Haumean family can be used to constrain the orbital conditions and physical nature of the collision that created the family, billions of years ago. Indeed, our results suggest that the formation of the Haumean collisional family most likely occurred after the bulk of Neptune's migration was complete, or even some time after the migration had completely ceased, although future work is needed to confirm this result.

**Keywords:** Kuiper belt: general – Kuiper belt objects: individual: Haumea – planets and satellites: dynamical evolution and stability – planets and satellites: individual: Neptune – minor planets, asteroids: general – methods: numerical

## 1 INTRODUCTION

Beyond the orbit of Neptune, the debris left over from the formation of our Solar system lies in cold storage. Two main reservoirs host that material. The Oort cloud (Oort 1950; Brasser, Duncan & Levison 2006) and the trans-Neptunian belt (also known as Edgeworth-Kuiper belt), populated by objects known as "Trans-Neptunian Objects" (hereafter, TNOs) (Edgeworth 1943, 1949; Kuiper 1951). The TNOs are often split into a number of classes, all of which contain objects that move on orbits with typical semi-major axes, $a$ < 2000-3000 AU. The study of these groups of TNOs can provide vital information about the origin and evolution of the Solar system and planet formation (Malhotra 1995; Horner, Mousis & Hersant 2007; Lykawka & Mukai 2007b; Lykawka & Mukai 2008; Morbidelli, Levison & Gomes 2008), and as such they are the targets of significant observational (Trujillo, Jewitt & Luu 2001; Bernstein et al. 2004; Muller et al. 2009, 2010; Lellouch et al. 2010; Lim et al. 2010) and theoretical work (Ida et al. 2000; Morbidelli, Emel'yanenko & Levison 2004; Lykawka et al. 2009, 2011; Lykawka & Horner 2010).

The classical TNOs orbit between roughly $a$ ~ 37 AU and 50 AU (the classical region), and typically occupy orbits that are dynamically stable on timescales comparable to the age of the Solar system (Levison & Duncan 1993; Holman & Wisdom 1993; Lykawka & Mukai 2005b). The structure of the classical population itself has turned out to be quite complicated, with objects ranging from the dynamically cold (as expected), to the dynamically excited "hot" population, with inclinations in excess of five or ten degrees. In addition, the trans-Neptunian population contains at least three further, unexpected dynamical classes of objects: resonant, scattered, and detached (Morbidelli & Brown 2004; Lykawka & Mukai 2005b; Elliot et al. 2005; Lykawka & Mukai 2007b, c; Gladman, Marsden & VanLaerhoven 2008).

Resonant TNOs inhabit a wide variety of mean-motion resonances[1] with Neptune, from the 1:1 (Trojans; Chiang & Lithwick, 2005; Zhou, Dvorak & Sun 2009; Lykawka et al., 2009; Sheppard & Trujillo 2010) to those beyond 50 AU (Chiang et al. 2003; Lykawka & Mukai 2007a, b; Gladman, Marsden & VanLaerhoven 2008). Importantly, the bulk of resonant TNOs move on orbits that are typically dynamically stable on Gyr timescales (Murray & Dermott 1999).

Other TNOs, by contrast, move on orbits that are significantly less dynamically stable. These objects are capable of experiencing significant gravitational scattering by Neptune (Duncan & Levison 1997; Gladman et al. 2002; Lykawka & Mukai 2006, 2007c). This sub-population can be further broken down into two components: "scattering" objects (recently perturbed objects on orbits currently strongly interacting with Neptune) and "scattered" objects (those more weakly perturbed by that planet). Scattering objects likely source a significant fraction of the Centaur population (objects moving on unstable orbits among the planets), which are, in turn, the main source of short period comets (Levison & Duncan 1997; Horner et al. 2003; Horner, Evans & Bailey 2004a, b; Volk & Malhotra 2008; Bailey & Malhotra 2009; Horner & Jones 2009). For simplicity, we call all such unstable trans-Neptunian objects the scattered TNOs.

Lastly, the detached TNOs move on orbits that resemble those of scattered TNOs, but are sufficiently detached from the gravitational influence of the giant planets that they do not suffer significant perturbations by them, even on timescales comparable to the age of the Solar system. This seems to be the case for objects moving on moderate or highly eccentric orbits with perihelia, $q$, greater than 40 AU (Gladman et al. 2002; Lykawka & Mukai 2007b; Gladman, Marsden & VanLaerhoven 2008; Lykawka & Mukai 2008).

In addition to the information that can be gleaned through detailed dynamical studies of TNOs and other minor bodies in the Solar system, the characterisation of the physical properties of these objects provides further precious clues to the origin and evolution of the planetesimal disk, and their ongoing evolution since that process came to an end (Horner, Mousis & Hersant 2007; Stansberry et al. 2008; Horner et al. 2008; Muller et al. 2010; Lellouch et al. 2010; Lim et al. 2010). A notable example of such work is the way in which theoretical and laboratory studies of collisions between rocky bodies have enhanced our understanding of the formation and evolution of the collisional families in the main asteroid belt (Nesvorny et al. 2002; Parker et al. 2008). By better understanding those families, we are then able to paint a better picture of the primordial population of objects in that region, both in terms of the original size/mass distribution, and the distribution of their chemical properties at the time they formed. Such studies also add constraints to our understanding of the internal structure of those bodies, post-collision satellite formation, and collisional physics (Gaffey et al. 1993; Benz & Asphaug 1999; Bendjoya & Zappala 2002; Michel, Benz & Richardson 2004b; Bottke et al. 2005). Indeed, the collisional fragmentation of the larger members of the asteroid belt is also well documented, with the presence of a number of notable asteroid collisional families having been traced back to collisions dates as recent as a few million years ago (Marzari et al. 1998; Nesvorny et al. 2002; Bottke, Vokrouhlicky & Nesvorny 2007).

[1] For brevity, 'resonance' will refer to external mean motion resonances with Neptune henceforth.

Given the number of objects orbiting in the trans-Neptunian region, it is reasonable to assume that the collisional evolution of that population has also continued throughout the entire lifetime of the Solar system. It is therefore reasonable to expect that collisional families will exist in the trans-Neptunian population, just as they do in the asteroid belt. This conclusion is supported by the presence of small satellites around a number of the larger TNOs (such as Pluto and Eris), which are believed to be evidence that those objects were once involved in significant collisions. To the best of our knowledge, Chiang (2002) represents the first attempt to search for collisional families within the trans-Neptunian belt, through the analysis of a large sample of known TNOs. However, that study was not conclusive, as a result of the difficulty of unambiguously separating family candidates from the background distribution of TNOs[2].

In 2007, Brown et al. overcame this problem by studying the orbital clustering of a group of TNOs in concert with their physical properties (determined through their observed spectra), thus providing appealing evidence for the first collisional family in the trans-Neptunian belt. At the time of discovery, the family consisted of the dwarf planet (136108) Haumea (formerly 2003 EL61) and five smaller TNOs. As of 25$^{th}$ November 2011, a total of eleven members of the family have been definitively identified (see Table 1)[3].

| Prov. Des. | $a$ (AU) | $e$ | $i$ (°) | $\Omega$ (°) | $\omega$ (°) | $M$(°) | $q$ (AU) | $\Delta v_{min}$ (m s$^{-1}$) | $H$ | $D$ (km) | Class |
|---|---|---|---|---|---|---|---|---|---|---|---|
| (136108) Haumea | 42.99 | 0.198 | 28.2 | 122.1 | 239.9 | 204.0 | 34.48 | - | 0.1 | 1500 | 12:7 |
| (145453) 2005 RR43 | 43.42 | 0.143 | 28.5 | 85.8 | 279.9 | 36.0 | 37.22 | 111.2 | 3.9 | 400 | Classical |
| (55636) 2002 TX300 | 43.50 | 0.126 | 25.8 | 324.5 | 342.6 | 63.1 | 38.03 | 107.5 | 3.2 | 350 | Classical |
| (120178) 2003 OP32 | 43.43 | 0.107 | 27.1 | 183.1 | 71.6 | 60.5 | 38.78 | 123.3 | 4.0 | 350 | Classical |
| (19308) 1996 TO66 | 43.50 | 0.116 | 27.4 | 355.2 | 242.5 | 126.7 | 38.46 | 24.2 | 4.4 | 300 | Classical |
| (24835) 1995 SM55 | 41.96 | 0.106 | 27.0 | 21.0 | 69.2 | 323.2 | 37.51 | 149.7 | 4.7 | 250 | Classical |
| 2005 CB79 | 43.17 | 0.139 | 28.7 | 112.9 | 92.5 | 310.1 | 37.16 | 96.7 | 5.0 | 250 | Classical |
| 2003 UZ117 | 44.36 | 0.134 | 27.4 | 204.6 | 245.2 | 333.3 | 38.41 | 66.8 | 5.4 | 200 | Classical |
| 2003 SQ317 | 42.90 | 0.085 | 28.5 | 176.3 | 192.5 | 357.2 | 39.24 | 148.0 | 6.5 | 100 | Classical |
| (86047) 1999 OY3 | 44.07 | 0.171 | 24.2 | 301.8 | 306.6 | 53.1 | 36.53 | 292.8 | 6.7 | 100 | Classical |
| 2009 YE7 | 44.57 | 0.138 | 29.1 | 141.5 | 100.3 | 174.1 | 38.43 | | 4.3 | | Classical |

**Table 1:** List of the currently known Haumea family members. The orbital elements (rounded off here for readability) and the absolute magnitude, $H$, of the objects were taken from the Asteroids Dynamic Site –

[2] This is a consequence of the fact that hypothetical collisional family objects in the trans-Neptunian belt will appear spread over large areas of semi-major axis (e.g., Ragozzine & Brown 2007).
[3] The identification of 2009 YE7 as a family member was announced (Trujillo, Sheppard & Schaller (2011) during the revision of this paper, and it was added to this table at that point for completeness. As such, it was not included in the construction of theoretical Haumean collisional families.

AstDyS[4] on the 30th September 2010. Here, *a* gives the semi-major axis, *e* the eccentricity, *i* the inclination of the orbit, *Ω* the longitude of the orbit's ascending node, *ω* the longitude of the object's perihelion, *M* the mean anomaly of the object on the on 30th September 2010, and *q* the perihelion distance. The orbits are described by their osculating elements at the current epoch. $\Delta v_{min}$ describes the minimum ejection velocity required for a fragment at the location of family formation to reach the orbit of the TNO in question, according to Ragozzine & Brown (2007) (with the values taken from that work). In addition, *D* represents the estimated diameter of the object (assuming a spherical shape) based on its *H* and assumed albedos of 0.75 (Haumea), 0.88 (2002 TX300) and 0.35 for all other family members (i.e., in agreement with the lower limits given in Rabinowitz et al. 2008). The dynamical class is given in the final column, where '12:7' means the 12:7 external mean-motion resonance with Neptune and 'Classical' stands for a region located approximately at 37-50 AU within the trans-Neptunian belt. The membership of 2009 YE7 in the Haumea family was identified during the revision of this paper, and that object was added to this table on 25th Nov 2011 for completeness. See main text for more details.

Haumea currently occupies a moderately eccentric ($e \sim 0.2$) and inclined ($i \sim 28°$) orbit at $a \sim 43$ AU. As a result of the orbit's eccentricity, Haumea reaches perihelion at approximately 34.5 AU, significantly closer to the orbit of Neptune than the nominal inner edge of the classical trans-Neptunian belt, at around 37 AU. As can be seen in Table 1, we confirmed that Haumea's orbit lies in the middle of the 12:7 resonance, which acts to stabilise the orbit (Lykawka & Mukai 2007b; Ragozzine & Brown 2007). Long term integrations of the orbital evolution of Haumea support this conclusion, with the object avoiding serious dynamical perturbation on long timescales.

Haumea is one of the largest TNOs known to date, with an estimated diameter of ~1500 km, mass of $4.1\text{-}4.3\text{x}10^{21}$ kg, and an unusually high albedo of 0.65~0.85 (Brown et al. 2005; Rabinowitz et al. 2006; Stansberry et al. 2008; Brown 2008). In addition to this extreme albedo, Haumea displays further unexpected and unusual properties, as described by Rabinowitz et al. (2006) and Lacerda, Jewitt & Peixinho (2008). Recently, radiometric fits to data taken using the *HERSCHEL* and *Spitzer* space telescopes suggest that the equivalent diameter of the object might be slightly smaller, at 1300 km, with an albedo of 0.70-0.75 (Lellouch et al. 2010). The obtained diameter is smaller than our estimated value in Table 1 because a larger value of $H \sim 0.4$ was used by Lellouch et al. However, the precise size adopted for Haumea in this work is unimportant for the modelling and conclusions of this work, as we explain in Section 2.

In addition, near infrared observations of the dwarf planet and the other TNOs in the Haumea family have yielded peculiar spectra that strongly suggest that their surfaces are simultaneously very rich in water ice and lacking compounds containing carbon (C-depleted), compared to more typical TNOs (Trujillo et al. 2007; Pinilla-Alonso et al. 2009; Pinilla-Alonso et al. 2007; Barkume, Brown & Schaller 2008; Pinilla-Alonso, Licandro & Lorenzi 2008; Schaller & Brown 2008; Rabinowitz et al. 2008). When one considers the orbital distribution of the objects in the Haumea family, it is clear that their orbits are clustered in the region of element space around $a \sim 42$ - 44.5 AU, $i \sim 24$ - 29° - a clustering that would be expected from objects with a common origin. However, aside from Haumea, the members move on orbits with significantly lower eccentricities, such that their perihelia are concentrated around $q \sim 37$ - 39 AU (see Fig. 1).

Further evidence for the Haumea collisional family comes from observations of Haumea's two satellites (Barkume, Brown & Schaller 2006; Fraser & Brown 2009; Ragozzine & Brown 2009), statistical analysis of the near infrared colours of several TNOs and those of the family members (Snodgrass et al. 2010), and the determination of the unusually high albedo (0.88) of the family member (55636) 2002 TX300 (Elliot et al. 2010). These results add weight to the conclusion that all these objects possess a common origin.

How populous is the intrinsic Haumea collisional family? Beyond the 11 currently known family members, it is likely that more will be discovered over the coming years. Wide-area surveys have been responsible for the discovery of the majority of the bright (large) TNOs, down to apparent R-band magnitude of ~21 (Schwamb et al. 2010; Sheppard et al. 2011). In addition, spectroscopic surveys using the HST have been used to identify water signatures in the spectra of more than 100 TNOs, including fainter objects. However, these surveys were unable to find new family members, aside from 1999 OY3 (Benecchi et al. 2011; Fraser, Brown & Schwamb 2010). More recently, Trujillo, Sheppard & Schaller (2011) performed a systematic survey with the specific goal of searching for water and methane ices on the surfaces of 51 TNOs, which resulted in the identification of just one new family member, 2009 YE7. In sum, when we consider the constraints and results from these dedicated surveys, it seems reasonable to assume that the majority of the largest members of the Haumean family have been already discovered. On the other hand, other studies suggest that TNOs that lack strong water ice features in their spectra could still be dynamically tied to the family, meaning that a number of new large members may await identification (Benecchi et al. 2011). Interestingly, new techniques for family identification based on statistical analysis of groups of objects compared to the background population have become available and do not require knowledge of surface properties such as colours or spectra (Marcus et al. 2011). This suggests that it is possible to identify both the currently known Haumean family and new members purely on theoretical grounds.

[4] http://hamilton.dm.unipi.it/

In the original collisional impact scenario, as suggested by Brown et al. (2007), Haumea is considered to be the largest undisrupted fragment remaining from the giant impact that created the family, whilst the other TNOs are thought to be fragments resulting from the ongoing collisional evolution of the family after that impact. However, that hypothesis has recently been challenged by the development of more sophisticated collisional scenarios, and as such, the details for the creation of Haumea's collisional family remain under debate[5] (Schlichting & Sari 2009; Leinhardt, Marcus & Stewart 2010). Despite the apparent uncertainty, however, the various scenarios all invoke the idea that the main family formation mechanism involves the collision of primitive bodies during the early Solar system and that the population of such bodies in the trans-Neptunian region was much greater at that time, in order that the collision which formed the family be plausible.

In this work, we focus on the long-term dynamical evolution of primordial Haumea collisional family members after the main event that created the family. To our knowledge, the closest theoretical work related to the orbital evolution of the family was carried out by Ragozzine & Brown (2007) and Levison et al. (2008). In the first work, the authors analysed in detail the circumstances of the collision that supposedly created the Haumea collisional family. They determined the collision location and the size of the instantaneous collisional cloud, aiming to constrain the most plausible family members among the known population of TNOs. Through analysis of the timescale required for objects to evolve from the location of the collision to orbits with eccentricities similar to that of Haumea within the 12:7 resonance, they determined that the family must have been formed at least 1 Gyr ago. In their 2008 work, Levison et al. constrained the orbital nature of the impactor and target bodies that created the Haumea collisional family. They suggested that the two bodies were in fact primordial scattered TNOs that collided within the classical region.

Although these studies looked at the initial evolution of the collisional family immediately after its formation, to date, no work has been carried out studying the long-term evolution of the family. Since the original collisional family has almost certainly undergone significant evolution since its formation around the birth of the Solar system, it is important to investigate that evolution by using numerical integrations. In this work, based upon reasonable assumptions of the initial, post-collision, distribution of objects within the primordial family, we aim to provide tentative answers to the following questions:

[5] However, the Haumean family may not be collisional in origin. See Ortiz et al. (2011) for an alternative scenario.

- How have the orbits of Haumea and its family members evolved over the age of the Solar system?
- What fraction of the initial population has survived to the current day, and how are those survivors currently distributed among the four main dynamical classes of TNOs[6]?
- How do those fragments that are either directly placed on, or acquire, highly unstable orbits diffuse through the scattered, Centaur and short period cometary populations?
- Which regions in element space yield the highest probability of finding new family members?

In addition to allowing us to better constrain the spatial distribution of members of the Haumea family at the current epoch, studies of the dynamics of such collisional families (both specifically the Haumea family, and other hypothetical families) can help provide fresh constraints on the importance of such collisions in shaping the orbital structure of the trans-Neptunian region. Future comparisons between observed family members and the theoretically predicted current distribution of objects in the family (taking into account their long-term, post-formation orbital evolution) will play an important role in determining which of the various scenarios for the origin of the family best fit its current distribution by constraining certain key parameters of the collision itself (such as the distribution of fragment ejection velocities).

## 2 MODELLING THE HAUMEA COLLISIONAL FAMILY

We modelled Haumea's primordial collision family by constructing clouds of fragments (theoretical families) at a time shortly after the collision had occurred. Our aim was to investigate the long-term orbital evolution of fragments with a given distribution of ejection velocities. In order to achieve this, we assigned ejection velocities to each fragments of the collision based upon the behaviour of a standard theoretical family with a given primordial size distribution. Those distributions were later used to obtain the initial conditions for our models of the orbital evolution of the theoretical families created. We justify this approach and describe the whole process in detail below.

First, the size distribution within these families was determined using the size of known members (Table 1) as a guide. This allowed us to place loose constraints on the size distribution of the family itself, despite the fact that the true distribution of family members of any given size is still unknown. Indeed, it is possible that objects of diameter a few hundred kilometres remain to be discovered, or to be associated with the family. Similarly, the well known observational bias against the discovery of smaller TNOs naturally limits the number of small family members that have been found to date, although numerical simulations of the formation of Haumea suggest that there should be many such bodies (Leinhardt, Marcus & Stewart 2010). In addition, it should also be noted that such objects are significantly harder to firmly tie to the Haumea family, since it is more challenging to obtain good quality spectra for fainter objects (Barucci et al. 2008). Indeed, only observations using the HST, or sophisticated theoretical techniques, could identify such faint candidates (See Section 1). Taking these concerns into account, we decided to only consider the largest members of the Haumea family (those objects with estimated $D$ > 200km) in order to determine a rough initial size distribution. This resulted in the three smallest members of the family being excluded from our analysis. Furthermore, a number of numerical simulations on the formation of collisional families have shown that the largest fragment is usually an outlier in the size distribution (Michel, Benz & Richardson 2004a; Leinhardt, Marcus & Stewart 2010). Given that Haumea is approximately four times the diameter of the next largest family member, and since we were interested in the size distribution of smaller objects, we also excluded Haumea. Once these cuts were made, the cumulative size distribution of the family was obtained according to

$$N(> D(km)) = KD^{-p}, (1)$$

[6] The four classes being the classical, resonant, scattered and detached groups, as described earlier in the introduction.

where $K$ is a constant set arbitrarily to equal $5x10^5$ (with units of $km^p$), $D$ is the object's diameter and $p$ is the decay exponent of the power law. Due to the complete lack of information about the properties of collisional families involving TNOs and the various observational biases discussed above, we opt to study the simplest case of a population distribution with a constant slope of $p = 2$, as guided by the six large members of the Haumea family (Fig. 2), and which is also, to first order, comparable to that inferred for several old asteroid collision families (Parker et al. 2008). Assuming a higher albedo for five of the six family members would shift the curve to smaller diameters, but would have only a minimal effect on its slope, $p$. Despite the difficulties discussed above, we believe our obtained distribution is an acceptable approximation to the size distribution of the fragments within Haumea's collisional family. Detailed simulations (such as carried out by Leinhardt, Marcus & Stewart 2010) of the formation of collisional families in the trans-Neptunian belt would be required in order to provide more realistic estimates and constraints on the size distributions and physical properties of such families and such modelling is beyond the scope of this work.

Second, based on the assumed cumulative size distribution and parameters associated with it (as detailed above), we created a standard population of 1600 objects. Each object within that population was given a representative diameter (in km), $D$, determined by the relation $D_n = (n/K)^{-\frac{1}{p}}$. The parameter $n$ was varied between 2 (following the assumption that the largest fragment in the family is Haumea) and 1601 (the smallest fragment). Next, the ejection velocity distribution of fragments was modelled to be $v_{eje} \propto m^{-\alpha}$, where $m$ is the fragment mass. Since the mass of a given family member is directly proportional to the cube of its diameter, this power law becomes the following relation

$$v_{eje} = QD^{-\beta}, \quad (2)$$

where $Q$ is a constant (with units of $m^{(1+\beta)}s^{-1}$). This results in the smallest particles having the greatest ejection velocities. Previous studies have shown $\alpha$ to lie approximately in the range 0 - 1/6, which corresponds to values of $\beta$ between ~0 and ~0.5 (Zappala et al. 2002). Since our knowledge of the exact nature of the family-forming collision is limited, we chose to consider two extreme examples within this range, with $\beta = 0.025$ and $\beta = 0.25$, respectively (hereafter "small" and "big"). These yield individual particle velocities that are very weakly, and strongly, dependent on the size of the particle, respectively. The values of $Q$ used to determine the ejection velocities for the "small" and "big" values of $\beta$ were 85 and 255, respectively. These values were arbitrarily chosen to approximately match the domain of minimum ejection velocities at size ~1 km (with the implicit understanding based upon the constraints posed by several studies, as discussed below).

When we apply Eq. 2 to our standard 1600-fragment cloud, it yields a suite of ejection velocities that follow a simple trend as a function of the diameter of the objects within the cloud, and the velocities obtained vary only slightly. However, the distributions of ejection velocities obtained in laboratory impact experiments (Holsapple et al. 2002; Giblin, Davis & Ryan 2004), through theoretical studies of collisional phenomena (Michel et al. 2002; Michel, Benz & Richardson 2004a, b; Jutzi et al. 2010; Leinhardt, Marcus & Stewart 2010), and through analyses of Lunar/Martian craters (Vickery 1986; Hirase, Nakamura & Michikami 2004) clearly show that fragments are ejected with a relatively wide range of velocities (see also Zappala et al. 2002 and references therein). Importantly, as discussed in several of the aforementioned studies, the obtained distributions often show a number of common features, such as large ratios of maximum to minimum ejection velocities and the mean ejection velocities of the bulk population being several times the velocities of the largest fragments. Such features can, for example, be seen in the skewed distributions of ejection velocities for theoretical asteroid families obtained by Michel et al. (2002). In addition, the apparent (in)dependence of ejection velocity upon fragment size can be clearly seen in a variety of studies[7]. This strengthens our case in testing two distinct values for beta, as described above.

Clearly, the distributions of ejection velocities obtained in the aforementioned studies are far more complex than that which is produced by our Eq. 2, so in an attempt to get qualitatively comparable distributions for our study of the Haumean family, we applied an extra random velocity increment to the velocities produced by Eq. 2. Following this procedure, the ejection velocities for each particle *n* were calculated following

$$v_{eje,n} = v_{eje}\left(1 + Fp(x)\right), \quad (3)$$

where $F$ is a control parameter (constant) and $p(x)$ gives the probability density function of the random velocity increments applied. We described $p(x)$ through use of Weibull distributions[8] and set $F$ = {17.1, 8.9, 31.1, 18.8, 45.1, 28.7} for Runs 1-6, respectively (see Table 2). These distributions were used solely as a mathematical tool to produce skewed ejection velocity distributions qualitatively comparable to those obtained in the literature. However, given the uncertainties involved (e.g., the variations observed within any one single model, such as that shown in fig. 5 of Michel et al. 2002, and the lack of available detail on the distributions from other such studies), other distributions with non-zero skewness may also be capable of producing acceptable velocity distributions[9]. In addition, we varied $F$ to adjust the distribution produced in each main run to obtain simultaneously the average ejection velocity of the distribution of interest (e.g., 300 m/s) and an appropriate ratio of maximum to minimum ejection velocities, as shown in Table 2.

Importantly, the ejection velocities obtained within our model distributions cover the values initially proposed in the scenario of Brown et al. (2007) (i.e., 140 and 400 m/s), and also those obtained by the analytical model of Schlichting & Sari (2009) (i.e., 120-190 m/s). In addition, the majority of our model ejection velocities also fall within the range of typical values obtained in the detailed numerical study performed by Leinhardt, Marcus & Stewart (2010) (so, for example, ~90% of our fragments had $v_{eje}$ < 300 m/s in the theoretical families described by our Runs 1, 1a-c, 4-100+, and 4-200+).

Despite the uncertainties and lack of knowledge about collisions between icy/rocky bodies in the trans-Neptunian belt, our set up clearly represents a more appropriate treatment of the plausible ejection velocities of fragments within the primordial Haumean family than previous studies which universally assumed a simplistic constant ejection velocity value for all fragments (Brown et al.

---

[7] Examples include fig. 2 of Zappala et al. (2002); fig. 5 of Michel et al. (2002); figs. 3 and 4 of Michel, Benz & Richardson (2004a); figs. 9 and 10 of Giblin, Davis & Ryan (2004); and figs. 11 and 16 of Jutzi et al. (2010).

[8] Random distributions obeying the probability density function, $p(x) = \frac{k}{\lambda}\left(\frac{x}{\lambda}\right)^{k-1} \exp^{-\left(\frac{x}{\lambda}\right)^{k}}$, set with shape parameter $k$ = 1.5 and scale parameter $\lambda$ = 0.1 for $x \geq 0$. The obtained values were constrained to remain between 0 and 0.4. A Weibull distribution (rather than, say, a simple Gaussian) was chosen due to its flexibility in producing a wide variety of data distributions, and its use is justified by the complexity and constraints of the ejection velocity distributions obtained in the literature, as discussed in the main text. Lastly, the values of F, $k$ and $\lambda$ were chosen for consistency with the desired parameters, as shown in Table 2.

[9] We note that only highly detailed simulations of collisions involving icy/rocky objects in the outer Solar system, taking into account their internal structure and physical properties, will allow us to finally determine which distribution is the most suitable to describe the fragment ejection velocities. The results of such simulations will also allow us to understand how the particular properties of the final distribution are related to the physical conditions of the collision, and the properties of the bodies involved. However, such highly detailed analysis is beyond the scope of this work.

2007; Ragozzine & Brown 2007; Levison et al. 2008). The ejection velocity distributions used for the six main runs of our simulations are illustrated in Fig. 3.

The initial ejection velocities of the test particles in our collisional families, as obtained by Eq. 3, were distributed isotropically about that of the collision location (i.e., the point determining the Keplerian vector velocity) from which the family originated, with the ejection direction of each individual particle being randomly assigned. We assumed the collision location in proper Keplerian elements to be ($a$, $e$, $i$, $\omega$, M) = (43.10, 0.118, 28.2, 270.8, 75.7) and set $\Omega = 0$, as determined and discussed in Ragozzine & Brown (2007) (see also Fig. 1). All members of the collisional families were treated as test particles that started at the same spatial location (that of the collision location) at the beginning of the simulations. Notice that no particular model of the formation of the Haumean family was assumed in this work, so that the details on the collision physics (e.g., target and impactor properties, re-accumulation processes, and so on) are beyond the scope of this work (Refer to Brown et al. 2007; Schlichting & Sari 2009; Leinhardt, Marcus & Stewart 2010 for such details).

Fifteen simulations were conducted in total to investigate the long-term orbital evolution of our theoretical representative collisional families. We varied a number of key parameters in the simulations, namely the number of massive bodies in the system, the slope of the power law given in Eq. 2 ($\beta$), and the initial ejection velocity distributions (representing the associated kinetic energies acquired by the fragments) (Table 2). We performed simulations of the Haumea family using 1600 particles under the gravitational influence of the four giant planets in each of the runs (Runs 1-6, Table 2), using the *Hybrid* integrator within the dynamics package MERCURY (Chambers 1999). In six additional simulations (Runs 1a-c and 3a-c), we explored the gravitational influence of Haumea, the largest fragments and the most massive TNOs currently known, Pluto and Eris, to see whether the influence of such bodies would result in significant differences in the orbital evolution of the theoretical families. In these particular runs, we included representative objects for Haumea and the other large family members, and the real objects for Pluto and Eris. For Haumea, we considered an object under the influence of the 12:7 resonance and close to the centre of the family cloud (before it acquired its current eccentric orbit), while for the other large family members we took orbits very similar to those observed today. Finally, in three other simulations, under the assumption that ejection velocities strongly depend on fragment size ($\beta = 0.25$), we investigated this dependence by following the evolution of fragments with a velocity distribution based on a minimum member size of 50 km, 100 km, and 200 km, using 1300 particles in each case (Runs 4-50+, 4-100+ and 4-200+, Table 2). These simulations concentrated solely on the evolution of the largest (and therefore most detectable) members of the Haumea family.

In all simulations the systems evolved over 4 Gyr to see how the orbital elements of the theoretical family fragments diffused over time. Particles were removed from the simulation when they collided with a massive body, or passed beyond 2000 AU from the Sun. The used time step was 243.5 days (0.67 yr). This time step is small enough to accurately model the orbits of TNOs (e.g., Morbidelli 2002). Unless specifically stated in the text, the outcomes of the simulations here are shown as instantaneous orbital elements. Wherever appropriate, we also used averaged orbital elements ("proper elements") to represent the outcomes from the simulations.

| Run | $\beta$ | Mean (median) $v_{eje}$ (m/s) | Ratio $v_{eje,max}$ / $v_{eje,min}$ | $N$ | Massive bodies | $f$ (%) |
|---|---|---|---|---|---|---|
| **1** | **small** | **200 (185)** | **6** | **1600** | **GPs** | **74.3** |
| **2** | **big** | **200 (188)** | **6** | **1600** | **GPs** | **74.8** |
| **3** | **small** | **300 (272)** | **10** | **1600** | **GPs** | **67.0** |
| **4** | **big** | **300 (268)** | **10** | **1600** | **GPs** | **66.4** |
| **5** | **small** | **400 (361)** | **14** | **1600** | **GPs** | **62.6** |
| **6** | **big** | **400 (352)** | **14** | **1600** | **GPs** | **64.5** |
| 1a | small | 200 (185) | 6 | 1600 | GPs + Ha | 72.2 |
| 1b | small | 200 (185) | 6 | 1600 | GPs + Ha + 4fm | 72.9 |
| 1c | small | 200 (185) | 6 | 1600 | GPs + Ha + 4fm + P&E | 70.0 |
| 3a | small | 300 (272) | 10 | 1600 | GPs + Ha | 65.2 |
| 3b | small | 300 (272) | 10 | 1600 | GPs + Ha + 4fm | 64.6 |
| 3c | small | 300 (272) | 10 | 1600 | GPs + Ha + 4fm + P&E | 63.4 |
| 4-200+ | big | 175 (157) | 8.5 | 1300 | GPs | 73.7 |
| 4-100+ | big | 202 (180) | 10 | 1300 | GPs | 72.5 |
| 4-50+ | big | 239 (214) | 11.5 | 1300 | GPs | 69.8 |

**Table 2:** Details of the fifteen 4 Gyr duration simulations performed in this study. The six main runs of our simulations are shown in bold, whilst subsidiary simulations are listed afterwards. The first number in the 'Run' column indicates the parameters of the main simulation considered. Runs 4-200+, 4-100+ and 4-50+ considered particles with ejection velocities determined such that all particles had effective diameters greater than 200 km, 100 km and 50 km, respectively. Runs 1a-c and 3a-c refer to special simulations that included the presence of massive trans-Neptunian objects and family members, in addition to the giant planets. $\beta$ is the exponent of the ejection velocity distribution, $v_{eje}$, as a function of size (Eq. 2), where 'small' and 'big' stand for $\beta = 0.025$ and $\beta = 0.25$, respectively. The mean (median) velocities and the approximate ratio of maximum to minimum velocities are given for each scenario considered (see Fig. 3). $N$ gives the number of particles modeled in the theoretical family cloud, whilst 'GPs' (the four giant planets), 'Ha' (Haumea), '4fm' (the four most massive Haumea family members after Haumea itself), and 'P&E' (Pluto and Eris) refer to the massive bodies included in the runs, respectively. Finally, $f$ gives the survival fraction of family members after 4 Gyr.

## 3 RESULTS

The majority of all objects created in our theoretical families survive within the trans-Neptunian belt for the full 4 Gyr of our simulations (see Table 2). In each case, the surviving members are wholly able to explain the observed orbital distribution of known Haumea family members in orbital element space ($a$-$e$-$i$). The features reproduced include the clustering of objects between $a \sim$ 42-44.5 AU, $e \sim$ 0.1-0.2 and $i \sim$ 24-28.5°. It should be noted that the results for scenarios with small values of $\beta$ are always essentially indistinguishable from those for scenarios with larger $\beta$. Since the precise value of $\beta$ is only important when discussing the variation of fragment ejection velocity as a function of fragment size (as discussed in Section 3.3), and given that, once the particles have been created, their size plays no role in the orbital integration over time, henceforth we will primarily present and discuss the results obtained from runs with small $\beta$ in our discussions, with the implicit understanding that the results of the equivalent run with large $\beta$ would be qualitatively the same. Hence, the results from Run 1 are reproduced by Run 2, Run 3 is reproduced by Run 4, and Run 5 is reproduced by Run 6.

### 3.1 ORBITAL EVOLUTION AND STABILITY OF COLLISIONAL FAMILIES

First, we found that the collisional families generated at the start of the simulations ($t$ = 0) evolve until they reach a steady state configuration, after approximately 1 Gyr. The subsequent evolution yielded no noticeable changes to the distribution in element space, except for the slow diffusion of objects that acquired unstable orbits during the last 3 Gyr of evolution. Indeed, we note that the obtained theoretical families after 4 Gyr resemble those in existence at the 1 Gyr mark (see Figs. 4-6). We illustrate the temporal evolution of the families with initial mean (median) ejection velocities, $v_{eje}$, of 200 (185), 300 (272) and 400 (361) m/s (Runs 1, 3 and 5) in Figs. 4, 5 and 6, respectively.

The velocity with which fragments are ejected in the collision that formed the Haumea family is actually a reasonable fraction of the typical Keplerian orbital velocity of TNOs, which results in the bulk of fragments being dispersed across a region spanning several AU. Indeed, the initial collisional families created with the lowest mean velocity (Runs 1 and 2) had extreme orbits separated by ~10 AU in semi-major axis. This extreme range increased to ~15 AU for the moderate case (Runs 3 and 4), and ~20 AU for the scenarios with greatest mean velocity (Runs 5 and 6) (see also Fig. 7). When the initial dispersion of family members is plotted in element space (Figs. 4-7), the outliers give the impression that the population is more widely distributed than is. In actuality, around 90% of the family members that survive the full 4 Gyr of the integrations are contained in the region $a \sim$ 40-47, 40-49 and 40-51 AU for Runs 1, 3 and 5, respectively. The dispersion of the fragments as a function of ejection velocity is illustrated in Fig. 8.

The fragments were also spread across a wide range of eccentricities (and hence perihelion distances, $q$) and inclinations in all runs, which allowed the majority of fragments that survived the 4 Gyr of integrations to populate wide areas of $a$-$e$-$i$ space, including the deeply stable regions of the belt, at $q > 40$ AU. A similar increasing dispersion of the inclinations and a tendency toward higher perihelia were also noted for the three scenarios represented by Runs 1, 3 and 5, respectively. It is worth noting that these final distributions for the obtained theoretical families strongly resemble their initial conditions at $a > 40$ AU and $q > 35$ AU. This has important implications for the Haumean collisional family, as we discuss in Section 4.1.

As the theoretical families generated at the start of the simulations contained ejecta fragments on varied orbits covering wide ranges of semi-major axis, eccentricity and inclination, it was essential to verify the stability of these objects and their possible mobility in element space over long time scales. Indeed, we have seen that the collisional clouds underwent more obvious dynamical evolution during the first 1 Gyr than the subsequent 3 Gyr (Figs. 4-6).

In general, a substantial fraction of the fragments were initially placed on orbits with $q < 40$ AU, orbits which, in principle, would be considered unstable on billion-year timescales (Holman & Wisdom 1993). However, perhaps surprisingly, the majority of these fragments survived for the full 4 Gyr of our study. This is a direct result of the high orbital inclinations of the fragments, which tend to substantially increase the stability of objects in this region at $a \sim$ 40-50(60) AU (see also Lykawka & Mukai 2005b). That said, a number of fragments acquired unstable orbits during the early stages of the simulations, as a result of close encounters with Neptune. This was particularly true of those objects that were placed on orbits with $q < 35$ AU. Indeed, a notable feature of the first 1 Gyr was the rapid depletion of the population of those small-$q$ objects at approximately $a < 42$ AU (Figs. 4-6).

At the end of the 4 Gyr simulations carried out in Runs 1 and 2, approximately 74-75% of the population created by the collision remained on orbits within the trans-Neptunian belt. The equivalent survival fractions for Runs 3 and 4 were of order 66-67%, while for Runs 5 and 6, they were of order 63-64% (see Table 2). It is not surprising that the simulations which involved the highest mean ejection velocity (and hence the most widely dispersed collisional fragments) displayed the lowest survival rates, but it is noteworthy that, in all cases, the majority of objects created in the collision survived for the age of the Solar system. The decay of the collisional clouds is illustrated by the results of Run 1, as plotted in Fig. 9. The behaviour within the other Runs was, qualitatively, very similar. Although the survivors represent a substantial fraction of the original modelled collisional family (at $t = 0$), and are spread over wide ranges of $a$-$e$-$i$ in all runs (Figs. 4-7), this spread is not uniform in element space. Finally, Fig. 10 indicates that even fragments ejected with high velocities can survive in the trans-Neptunian belt, provided that such objects acquire stable orbits (e.g., with $q > 35$ AU) after the creation of the family.

To better understand the distribution of the fragments, we computed the number density of family members for the three scenarios at the completion of the simulations at 4 Gyr, finding in each case that the surviving fragments were concentrated around the location of the family-generating impact ($a \sim 43.35$, $e \sim 0.126$ and $i \sim 27.7°$ in osculating elements) (see Fig. 11). The currently known members of the Haumean collisional family are concentrated around $a$ = 42-44.5 AU and $e$ = 0.1-0.2, thus lying in the region of highest number density obtained from our simulations. These density plots can be used to infer the regions in which undiscovered members of the Haumean family are most likely to reside, and therefore show where we believe such objects are most likely to be identified in the future.

## 3.2 DEPENDENCE OF THE EVOLUTION OF THE HAUMEA FAMILY ON THE GRAVITATIONAL INFLUENCE OF MASSIVE TRANS-NEPTUNIAN OBJECTS

How would the results of our study change if the gravitational influence of Haumea, the most massive family members, or the other dwarf planets, such as Pluto and Eris, were included in these calculations? As detailed in Section 2 and Table 2, we performed six subsidiary runs to investigate the influence of those massive TNOs on the long-term evolution of the theoretical families. The outcomes of Run 3 for four particular cases are illustrated in Fig. 12. The first of those scenarios follows the evolution of the family members under the gravitational influence of just the four giant planets. In the second, Haumea itself is treated as a massive body, and can influence the test particles being simulated in the same way that the giant planets do. In the third scenario, the four most massive Haumea family members are also treated as massive particles, in addition to Haumea itself, and the four giant planets. Finally, in the fourth scenario, the two most massive objects known in the trans-Neptunian region, Eris and Pluto, are added to the four giant planets, Haumea, and the four most massive fragments, meaning that a total of eleven massive bodies are considered in that run.

As can be seen in Fig. 12, which shows the end-of-simulation distribution of family members in *a*-*e* space for Runs 3 and 3a-c, the final distributions of surviving family members obtained from these tests are indistinguishable. The same result holds for the effect of such bodies on the final inclination distribution of family members and when we consider the results of Runs 1 and 1a-c. We therefore conclude that the gravitational influence of massive family members and other massive TNOs on the evolution of the Haumean collisional family is negligible. However, as can be seen in Table 2, the simulations which followed the evolution of the families under the influence of massive family members and TNOs did show a small, but significant, increase in the number of particles that were ejected over the course of the simulations (of order 2 - 4%, depending on the massive bodies included). In other words, although the presence of massive bodies did not appreciably alter the distribution of the surviving members at the end of the simulations, it did act to slightly increase the ease with which members could escape from the trans-Neptunian belt to dynamically unstable orbits, and hence be removed from the simulation. This is a tantalising hint that massive members of the trans-Neptunian population might play some role in influencing the rate at which TNOs are injected to the Centaur population (Horner et al. 2003; Horner, Evans & Bailey 2004a, b; Horner & Lykawka 2010a, b). However, in this case, the observed extra depletion levels are so small that we can conclude that the omission of massive family members and TNOs from our core simulations has not significantly compromised our conclusions.

## 3.3 COLLISIONAL FAMILY FRAGMENTS AND DYNAMICAL CLASSES IN THE TRANS-NEPTUNIAN BELT

As noted in Section 1, Haumea is a resonant TNO currently locked in the dynamically weak[10] 12:7 resonance located within the classical region of the trans-Neptunian belt. Despite the influence of the nearby 5:3, 7:4 and 12:7 resonances, the ten other members of the Haumea family are non-resonant, as determined from their best-fit orbits (Table 2). We were also unable to confirm that 1996 TO66 is trapped in the weak *i*-type 19:11 resonance (Lykawka & Mukai 2007b). In all simulations of theoretical families performed in this work, the majority of fragments remained on orbits within the classical region after 4 Gyr. More specifically, because of the high inclinations ($i > 20°$) acquired by these bodies, they would be classified as hot classical objects were they discovered today. It is worth noting that none of the fragments studied was able to evolve into the cold component ($i < 5$-$10°$) of the classical region in any of our simulations.

[10] Objects locked in weak resonances in general experience small eccentricity/inclination changes over very long timescales (Gyr). These resonances also have less ability to capture and retain TNOs than strong resonances (such as the Neptunian 3:2). See Gallardo (2006) and Lykawka & Mukai (2007a, c) for more details about resonance strength and stickiness.

On the other hand, a number of fragments were captured into distinct resonances across the trans-Neptunian belt within just a few million years of the start of our simulations. A fraction of such resonant bodies was even able to survive locked in resonance over the age of the Solar system (Figs. 4-7). In particular, the 3:2 ($a$ = 39.4 AU), 12:7 ($a$ = 43.1 AU), 7:4 ($a$ = 43.7 AU) and, to a lesser extent, the 2:1 ($a$ = 47.8 AU) resonances proved the most efficient at capturing collisional fragments. After examining the role these resonances can play influencing the long-term dynamical evolution of such objects, the 5:3 and 7:4 resonances were observed to play an important role in allowing higher mobility of ejecta fragments in eccentricity and inclination space. These results are in agreement with earlier studies that examined in detail the dynamics of trans-Neptunian resonances (Malhotra 1996; Nesvorny & Roig 2001; Lykawka & Mukai 2005a, 2006, 2007b; Chiang et al. 2007).

In addition to those particles evolving as classical and resonant objects, other fragments, particularly those that had $q$ < 37 AU at the start of the simulations, acquired orbits typical of scattered TNOs after suffering gravitational scattering by the giant planets. Several of these objects then evolved onto orbits within the Centaur and short period cometary populations, a process that occurred most frequently during the first 1 Gyr of evolution (Figs. 4-6), whilst a population of less stable objects remained. Nevertheless, this process likely continues until present time, albeit with a very small influx rate of fragments to the Centaur population (Fig. 9).

Finally, a number of fragments were placed on moderately eccentric orbits with $q$ > 37 AU, a fraction of which developed both unstable (e.g., scattered) and stable orbits. The fragments on stable orbits were represented mainly by classical objects within $a$ ~ 50 AU. The maximum perihelia observed for the stable fragments were approximately 42-43 AU. However, because there is no clear boundary between the outer classical region and the detached region, several fragments could be considered representative of detached populations, especially those with $a$ > 45-50 AU and $q$ > 40 AU (see Lykawka & Mukai 2007b; Gladman, Marsden & VanLaerhoven 2008).

In sum, the fragments of our theoretical families were able to populate each of the four main dynamical classes in the trans-Neptunian belt, with the surviving population after 4 Gyr concentrated within the classical and detached populations (approximately 90%). The fraction of fragments moving on stable resonant orbits summed no more than 10% of the final population (e.g., only ~1% of the objects were locked in the 7:4 resonance in each of the main runs). The lack of observed Haumea family members in resonances is consistent with this picture. Finally, we estimate that no more than 1% of the fragments were found moving on orbits typical of the scattered population. This supports the idea that the Haumea collisional family is older than 1 Gyr (Ragozzine & Brown 2007).

### 3.4 DEPENDENCE OF THE FINAL FAMILY DISTRIBUTION ON THE SIZE DISTRIBUTION OF COLLISIONAL FAMILY FRAGMENTS

How would the results change if the ejection velocity of fragments varied significantly as a function of fragment size? If the ejection velocity of fragments varied significantly as a function of fragment size, would such a scenario lead to distinct spatial distributions of large fragments and their smaller brethren? To address this question, we performed three further special runs based on the ejection velocity distributions that were strongly tied to fragment size ($\beta$ = 0.25) (see Section 2 for details). These runs can essentially be considered supplementary calculations to our main Run 4, which used the same initial parameters, but whose ejection velocities were constructed considering ejecta of all sizes. In order to examine the dispersal of objects with diameter below 50 km, 1450 such objects were analysed from Run 4, allowing direct comparison to the other cases. It is important to remember, here, that because the real physical size of the particles was not taken into account in the simulations (i.e., use of massless particles), the ejection velocities were the essential parameter that dictated the outcomes of the collisional families. In short, the results obtained from the orbital dispersion of the particles (as determined by their sized-dependent initial ejection velocities) were used as a proxy to infer the fate of fragments as a function of size.

First, we found that the larger fragments tended to remain less dispersed than their smaller counterparts at the end of our simulations. This tendency was also observed for the other main runs that invoked a large value for $\beta$ (Runs 2 and 6), despite the small number statistics of large fragments in these two particular calculations. This is not surprising, as the ejection velocities associated with particular size ranges ($D$ > 200 km, $D$ > 100 km and $D$ > 50 km) were confined to narrower ranges of smaller ejection velocities that was the case for the size-independent sample discussed above (Fig. 3). With this understanding in mind, Fig. 13 illustrates the obtained distributions of fragments as a function of their size from Runs 4, 4-200+, 4-100+ and 4-50+. If the ejection velocity of fragments truly depends strongly on their sizes, the relatively large number of test particles used in these simulations (with each following the evolution of 1300 objects of an appropriate size) allows one to readily see the most likely regions of element space in which the subpopulations within particular size ranges would reside after 4 Gyr. The different degrees of dispersion for the fragments reflect the spectrum of ejecta velocities considered in each run, and so we note that the distributions obtained from Runs 2 ($\langle v_{eje} \rangle$ = 200 m/s) and 6 ($\langle v_{eje} \rangle$ = 400 m/s) exhibit dispersions that are smaller and greater, respectively, than those presented in Fig. 13. In general, as a result of wide ranges of ejection velocities acquired by the theoretical smaller fragments, this increased dispersion resulted in these particular populations suffering greater depletion through the course of the simulations than the "larger" test particles that concentrated on smaller initial ejection velocities. Explicitly, in the case of our Run 4, and the special variants detailed above, the survival fractions for the two extreme size ranges after 4 Gyr were 73.7% ($D$ > 200 km) and 66.7% ($D$ < 50 km) (Table 2).

In sum, if collisional fragments are ejected with velocities inversely proportional to their sizes, then we can expect the largest fragments to survive more tightly clustered around the location of the family-generating impact, whilst smaller fragments (e.g., with diameters of tens of km or less) are expected to disperse into a significantly wider variety of orbits after billions of years of dynamical evolution. In this sense, the currently known Haumean collisional family members seem to follow this tendency: they are larger than about 100 km and are highly clustered in orbital element phase space. Following this logic, the identification of the smaller members of the Haumean family will be necessary before conclusions can be drawn on whether the impact fragments were dispersed initially on velocities that were related to their sizes.

## 4 DISCUSSION

In this section, we assume that the collisional and orbital evolutionary models described in this work are a fair representation of the evolution of the real primordial Haumean collisional family, and take the outcomes from the main simulations discussed in Section 3 as "true" possible distributions for the intrinsic family at current time. We also recall that the impact that created the family was modelled with a location set at $a \sim 43.35$ AU, $e \sim 0.126$ and $i \sim 27.7°$ ($q \sim 37.9$ AU) (see Sections 2 and 3 for details).

### 4.1 GENERAL TRENDS AND PREDICTIONS FOR THE HAUMEAN COLLISIONAL FAMILY

Our results suggest that the Haumean collisional family could be distributed within relatively clustered regions of the trans-Neptunian belt, if the kinetic energies acquired by the ejecta fragments were such that the mean ejecta velocity was ~200 m/s (Fig. 4). Alternatively, if the mean ejecta velocity was ~300 m/s (or even as high as ~400 m/s; Figs. 5 and 6), then the family could be currently distributed over much wider regions of $a$-$e$-$i$ space. Nevertheless, we note that in all scenarios considered (Table 2) the family members would most likely be concentrated in non-resonant orbits close to the collision location (Fig. 11). Indeed, the majority of family members (90% of the fragments) appear concentrated at $a$ = 40-47, 40-49 or 40-51 AU for the three main scenarios explored.

In addition, if the ejection velocities of the fragments that make up the Haumean family were dependent on the size of the fragments in question, with the largest fragments having the smallest mean ejection velocities, and the small fragments the highest mean velocities, then we can expect the largest members of the Haumea collisional family to be clustered relatively tightly around the collision location, with the smaller family members being dispersed on orbits covering wider regions of orbital element space around that location. This secondary effect would be overlain on that described above (i.e., the scenarios in which no dependence on fragment size was considered), with the overall distributions being described in that manner, and a size-dependent dispersion being apparent within that distribution. How, then, would the results change if the derived cumulative size distribution curve (Fig. 2) was obtained for family members with higher assumed albedos? As explained in Section 2, the curve would shift to smaller sizes, and so the fraction of ejecta incorporated into the smaller members of the family would increase. This would most likely result in slightly smaller survival fractions of family members, since smaller fragments would be more likely to acquire unstable orbits. However, we believe that this effect would be quite small, and thus not change the main results.

At any given time of our simulations, the fragments from the theoretical families were dispersed sufficiently to cover the entire region of element space which contains the eleven currently known members of the Haumean family (including Haumea itself), and beyond. Therefore, in principle, the impact that created the family as described herein could have emplaced even the more dynamically excited family members (such as Haumea and 1999 OY3) on their current orbits. Such injection provides an alternative mechanism to explain the high orbital eccentricities of these objects to the typically invoked long-term excitation by nearby resonances (the 12:7 for Haumea and 7:4 for 1999 OY3).

It is interesting, also, to examine the orbital evolution of those fragments which survived the full 4 Gyr of our simulations. Those objects can essentially be broken into two main groups – non-resonant objects, and those trapped in resonances. In the case of the non-resonant survivors, the great majority displayed little or no dynamical evolution over the course of the simulations, with their initial and final orbital elements being almost indistinguishable. By contrast, those objects which were captured into resonance displayed behaviour typical of resonant TNOs, with both eccentricity and inclinations varying while the semi-major axis of the orbit was constrained by the resonance. In fact, a direct comparison between the distributions of fragments at times $t$ = 0 and $t$ = 4 Gyr in Figs. 4-7 allows one to conclude that it is possible to use the current dispersion of the Haumean family in $a$-$e$-$i$ space to probe the initial properties of the collision (particularly the kinetic energy transferred to the fragments) that formed the family billions of years ago.

When we take into account the gravitational influence of the most massive TNOs in our simulations, a slight reduction in the survival fraction of family members was noted. Recalling the uncertainties and limited variation of initial conditions in our model, it is fair to conclude that the current day Haumean family should represent approximately 60-75% of the primordial family that was created during the early days of the Solar system. The remaining fragments (some 25-40% of those created) will have been lost from the family by dynamical evolution onto unstable orbits, with the eventual fate of the majority of such objects being ejection from the Solar system as a result of a close encounter with a giant planet. Some fraction of the fragments which left the trans-Neptunian region will certainly have become short period comets. Some might have experienced capture to pseudo-stable populations (such as temporary planetary Trojan orbits; e.g., Horner & Evans 2006), or even the irregular satellite populations of the giant planets (Jewitt & Haghighipour 2007).

Although our results seem to shed some light on the evolution of the Haumean family, the small number of family members currently known, and the difficulties inherent in the identification of new family members prevent us from performing more detailed comparisons of theoretical results with observations. On the other hand, as discussed in Section 1, if the currently known family members represent the intrinsic core of the family, their small spread in orbital elements would argue against scenarios in which the fragments occupy wide areas in element space. Thus, this would favour Run 1 as the best fit to the true Haumean family of the various scenarios considered in this work. Future observational work will identify more members of the Haumean family (e.g., Pan-STARRS and the LSST; Trujillo 2008 and references therein), hopefully uncovering members with diameters of tens of kilometres, or even smaller), and providing detailed information on the spread of their orbits, together with any apparent variation in their distribution as a function of size. This will allow the "true" boundaries of the family in orbital element space to be determined. Such data will allow the nature of the collision which formed the family to be much more rigorously determined, with the precise distribution of the family allowing the determination of factors such as the collisional energies involved, the dependence of ejecta velocity on fragment size, and perhaps even the physical properties of the impactor and target.

## 4.2 ORIGIN OF THE HAUMEA COLLISIONAL FAMILY

A number of studies have proposed scenarios for the creation of the Haumean collisional family, and the creation of the Haumea system itself (Brown et al. 2007; Schlichting & Sari 2009; Leinhardt, Marcus & Stewart 2010; Ortiz et al. 2011). Little attention has been given, however, to the orbital evolution of the family after the initial collision event, and the possible implications of the long-term behaviour of the family on our understanding of the wider study of the outer Solar system. For instance, precisely when the family was created during the early Solar system? Did the family-forming event occur before, during or after the large-scale migration of Neptune and other giant planets through the planetesimal disk? (as described by e.g., Levison et al. 2007 and references therein). Did the Haumean family really originate from a collision? (see an alternative model proposed by Ortiz et al. 2011)

One piece of evidence that points to the formation of the Haumean collisional family being early in the life of the Solar system is the moderately high eccentricity (0.2) of Haumea's orbit, which may be the result of long-term dynamical manipulation of the object while within the 12:7 resonance. The timescales for exciting orbits from an initial location near the family forming impact to that of Haumea are on the order of billions of years (Ragozzine & Brown 2007). In each run of our calculations, we typically found 5-10 fragments captured in the 12:7 resonance after 4 Gyr, one fragment of which had an eccentricity of order $e \sim 0.2$. Following the logic outlined above, Levison et al. (2008) suggested that the collision of two primordial scattered TNOs could have happened before Neptune finished its migration to its current orbit at 30.1 AU, happening, in other words, whilst the planet was still migrating.

If such a giant impact occurred when Neptune was located around its current orbit (as modelled in this work), then we expect that Haumea's orbit underwent gentle dynamical evolution until it happened be captured in the 12:7 resonance. At the same time, a small fraction of the total population of fragments were captured in a number of the web of resonances located between the 3:2 (inner edge) and 2:1 (outer edge) resonances (Figs. 4-7). The greatest captured populations are likely to reside in resonances located within a few AU form the collision location (such as the 8:5, 5:3, 12:7, 7:4, 9:5, 11:6, and potentially even weaker resonances in that region). We estimate that approximately 4-6% of the fragments will have survived locked in to those particular resonances, with the total population of resonant fragments likely making up less than 10% of the overall family.

Alternatively, if the family instead formed when Neptune was located at ~25-27 AU during its outward migration, as typically proposed by the model of Levison et al. (see also Morbidelli, Levison & Gomes 2008), then Neptunian sweeping resonances likely passed through the region containing the Haumea family in semi-major axis space, ~42-44.5 AU (although this region may actually be wider, as shown elsewhere in this work). The strongest sweeping resonances to affect the family, and their primordial initial locations, would be the 7:4 ($a_0$ ~ 36-39 AU) and 2:1 ($a_0$ ~ 40-43 AU), respectively. Because the fragments would possess a broad initial range of eccentricities, as indicated by the evolution of the fragments within 0-100 Myr (see Figs. 4-6), both resonances would have comparable probabilities of capturing a few tens of percent of the family members as they swept past it (Lykawka & Mukai 2007a, c for more details). In this case, in stark contrast to the scenario discussed above, it is likely that the fraction of Haumea family members currently trapped in resonances would be significantly larger than the 10% upper limit discussed above, and that these members will show no local preference for resonance occupancy around 41-45 AU. If the family fragments were initially distributed across a region as wide as proposed in this paper ($a$ = 40-49 ± 2 AU) after the family's birth, but before Neptune started its migration from the proposed ~25-27 AU, then it may be possible that more distant Neptunian resonances such as the 7:3 ($a_0$ ~ 44-47.5 AU) and 5:2 ($a_0$ ~ 46-50 AU) may have captured a significant fraction of those fragments during the subsequent planetary migration.

One potential problem with the idea that Neptune underwent significant migration after the formation of the Haumean family, however, is that it seems somewhat unlikely that an object such as Haumea would be preferentially captured by the weak 12:7 resonance, rather than one of the stronger resonances which swept the area (such as the 7:4 or 2:1), with much higher capture probabilities. Beyond this, the fact that the other nine members of the family are not trapped within any of the Neptunian resonances, particularly the 7:4 resonance, is suggestive of the fact that the family-forming impact occurred after the cessation of Neptunian migration, so that the 7:4 and other resonances did not have the opportunity to sweep through the family and capture a substantial number of fragments after the giant impact. Moreover, even in particular scenarios with substantial migration, it seems virtually impossible that the Haumea family was created elsewhere and that later the entire cloud was transported via resonant processes to its present location.

In any case, the lack of other resonant family members, coupled with Haumea's slow eccentricity excitation within the 12:7 resonance (1999 OY3 may also have suffered similar eccentricity excitation by the influence of the 7:4 resonance; Ragozzine & Brown 2007), seems to favour the formation of the Haumea family at a time when Neptune had already approached its current orbit, billions of years ago. This is in-line with the findings of Leinhardt, Marcus & Stewart (2010) in their model of the formation of the Haumean family. However, given the small number of members known, this could equally be the result of observation biases (preferential searching of the non-resonant population, for example), or simply bad luck in the detection of members. Again, the expected growth in the number of known family members over the coming years should help to shed light on which model best represents the formation of the family (pre- vs. post-migration).

### 4.3 THE ROLE OF COLLISIONAL FAMILIES IN THE TRANS-NEPTUNIAN REGION

Our results suggest that any event capable of producing a collisional family such as Haumea's will also spread fragments over wide ranges of *a-e-i* space. This spread can be substantial ($\Delta a$ > 10 AU) even for the most conservative low-energy impacts considered in this work.

At this point, we remind the reader that major collisions are thought to have played a significant role in the formation and evolution of the Solar system, in particular during its early stages, at which point the planetesimal disk was likely populated by at least two orders of magnitude more objects than is currently the case (Kenyon et al. 2008). Outstanding examples include: the accretion of giant planet cores (Cameron 1975; Pollack et al. 1996; Goldreich, Lithwick & Sari 2004), terrestrial planet formation and their subsequent shaping by giant impacts (Benz, Slattery & Cameron 1986; Benz et al. 2007; Davies 2008; Andrews-Hanna, Zuber & Banerdt 2008; Raymond et al. 2009), and a extensive gamut of outcomes during the collisional evolution experienced by the smaller members of the Solar system menagerie, as detailed in Section 1. It therefore seems certain that giant collisions such as that which created the Haumean family were the rule, rather than being rare, stochastic events, in the outer Solar system. We therefore expect that such impacts will have created many other collisional families within the trans-Neptunian belt, with the Haumean family merely being the first of many that will be identified in coming years.

Future studies may well identify collisional families associated with the origin of the satellite systems around the largest TNOs, such as the Pluto-Charon and Eris-Dysnomia systems. In this scenario, if fragments resulting from such giant impacts carried kinetic energies comparable to those used in this work, then these objects must have spread over wide areas in element space (*a*, *e*, *i*) beyond Neptune. In support of this hypothesis, in modeling the collisional origin of Haumea's family, Schlichting & Sari (2009) suggested the existence of ~30 collisional families originating from a population of ~520 km-sized progenitors ($D_P$) in the belt. Moreover, Marcus et al. (2011) developed a detailed theoretical model for the identification of hypothetical collisional families in the trans-Neptunian belt. They found that at least one collisional family for $D_P > 400$ km and ~20 families for $D_P \sim 300$ km should exist. The less energetic collisions also probably produced several "small scale" collisional families with fewer fragments and less dispersion. Since such collisions were more frequent than those associated with giant impacts (such as those invoked for Haumea, Pluto, etc.), the contribution of small collisional families may be considered important. Marcus et al. (2011) found that these small families may appear more clustered in element space and be more difficult to identify, even if such families are more frequent, when compared to large scale families ($D_P > 400$ km).

In conclusion, collisional families likely played an important role in shaping the orbital structure of the trans-Neptunian belt. One might even speculate that most TNOs acquired their orbits mainly from such collisional processes, in addition to the distant gravitational perturbations of planets and other massive bodies. A similar discussion on the importance of collisions on the orbital evolution of TNOs can be found in Levison et al. (2008). However, the likely ubiquitous existence of all these collisional families in the outer Solar system also poses the problem of how to uniquely identify such families, since their orbits will appear indistinguishable from the general background of those TNOs not related to families. Under the condition that the background population is well characterized, Marcus et al. (2011) propose a solution to this problem by identifying collisional families as statistical over-densities, greater than the expected fluctuations in the background population. Another promising technique consists of backwards integration to identify families through the clustering of objects' angular elements. They also suggest that dynamical evidence alone may be sufficient to determine which TNOs belong to which family. However, given the difficulties in uniquely identifying clumps of TNOs from the background with the current scarcity of observations, collaboration between theorists and observers will be vital, with a combination of spectral evidence, physical properties, and dynamical results coming together to help categorise the various families in the population beyond the orbit of Neptune.

### 4.4 A NOTE ON THE CONTRIBUTION OF COLLISIONAL FAMILIES TO THE COMETARY POPULATION AND THE POPULATION OF NEAR-EARTH OBJECTS

It has been suggested that a group of short period comets that display significant depletion of carbon might have their origin in those members of the Haumean family that are moving on dynamically unstable orbits (A'Hearn et al. 1995; Pinilla-Alonso et al. 2007, 2009; Pinilla-Alonso, Licandro & Lorenzi 2008). Given the general friability of cometary bodies, and the presence of large objects in the Centaur population (with the largest members having diameters of hundreds of kilometres), it is feasible that a fragmentational cascade as a large Haumean member moved inwards could source a large number of comets – and so this idea does not necessarily require an unfeasibly large unstable Haumean population to supply the observed comets.

Unfortunately, the data output time step used in our simulations prevents us from performing a detailed analysis of the dynamical transfer of theoretical family escapees onto cometary orbits. However, we note that lost members of the family regularly evolved onto typical Centaur-like orbits, with $i < 40°$, a fraction of which will eventually evolve to Jupiter-family comets (~10-30%, Levison & Duncan 1997; Horner et al. 2003; Horner, Evans & Bailey 2004a,b). On the other hand, because the Haumean family was likely created ~4 Gyr ago, the current supply of unstable family members should be very small. Indeed, we note that only ~1.9-2.5% of the family members were lost during the last 1 Gyr, in all simulations. Thus, taking optimistically the largest values above and assuming a total initial population of 1 million fragments with cometary sizes (say, between a hundred metres and <10 km) based on the same size distribution of Section 2, we estimate the injection of a fresh comet every 130,000 years. Since this timescale is somewhat longer than the typical dynamical lifetime of short period comets with estimated lifetimes typically of order $10^4$-$10^5$ years (Levison & Duncan 1994; Horner et al. 2003; Horner, Evans & Bailey 2004a,b), this would suggest that statistically the population of cometary-sized collisional family fragments cannot, at the present time, explain the origin of observed C-depleted comets. This conclusion remains valid even if the initial population of such fragments were 10 million objects (say, if a steeper slope was adopted for the size distribution, instead of -2, in Section 2), although we caution that the fragmentation of a larger escaped fragment as it evolved through the Centaur region would of course create a significant population of smaller cometary bodies, which would evolve independently. Therefore, there remains the possibility that at least some of the C-depleted comets are genetically linked to a large progenitor that fragmented after leaving the trans-Neptunian belt.

The transfer of material from the Haumean family to Jupiter-family comet orbits means that some of the debris created in that collision will have impacted on each of the planetary bodies in the Solar system. Given that a small number of fragments acquired orbits typical of Centaurs and comets over the last 1 Gyr of the simulations, presumably the fraction of those that acquired near-Earth orbits ($q < 1.5$ AU) or that collided with a terrestrial planet was negligible. Nevertheless, despite the limitations of data output resolution in our simulations, we estimate that the flux of unstable fragments acquiring cometary orbits and $q < 1.5$ AU is at least an order of magnitude higher during the first hundred Myr than that found during the last 1 Gyr of orbital evolution. This increased flux is illustrated on a long timer-scale by the relatively rapid evolution of the unstable fragments during the first 1 Gyr (as shown in Figs. 4-6).

## 5 SUMMARY OF CONCLUSIONS AND FUTURE WORK

In this work, we modeled the long-term orbital evolution of Haumea and its associated collisional family of fragments by constructing clouds of objects that were created by the giant collision thought to have created the family, billions of years ago (theoretical families). Through the course of this work, we examined the role of a number of key factors in the evolution of these theoretical families. First, the role played by the amount of kinetic energy deposited in the fragments by the initial collision was considered, with "slow", "medium" and "fast" scenarios being represented by ejecta distributions with mean ejection velocities of 200, 300 and 400 m/s, respectively. Secondly, we examined the gravitational influence of the most massive objects in the trans-Neptunian region (Pluto and Eris), together with the influence of the five largest known members of the Haumean family (including Haumea itself), on the long-term evolution of the family. Finally, based on the ejection velocity distributions assigned to larger and smaller fragments within a given family, we also examined the importance of fragment size on the eventual distribution of fragments, to see whether one would expect any significant clustering of larger bodies over their smaller brethren. This was achieved by comparing the effect of two different scenarios for ejection velocities – one in which the ejection velocity of a given fragment was only very loosely tied to its size, and one in which the ejection velocity was a strong function of fragment size (see Eq. 2 and Fig. 3). By far the most important of these three considerations turned out to be the first, the mean ejection velocity – with the second and third aspects studied having only minor effects in comparison. As such, this means we can describe the main results of this work in terms of three main scenarios.

Based on the results of our simulations, the main conclusions, implications and predictions (with the implicit assumption that this model correctly describes the evolution of Haumea's family) are summarized below. In this summary, the term "theoretical family fragments" refers to the objects that survived in the Solar system after 4 Gyr of orbital evolution.

- Even when there is significant variation in the key parameters of our simulations, we can accurately reproduce the orbital distribution of the currently known members of the Haumean collisional family the trans-Neptunian belt (Figs. 4-7). Our results do suggest, however, that the family occupies a wider region of the trans-Neptunian realm than currently constrained by observations.

- The theoretical family fragments are spread over a wide range of orbital elements (semi-major axes, eccentricities, and inclinations) in the trans-Neptunian belt (Figs. 4-7 and 12-13). The great majority of these fragments were distributed as follows (see also Fig. 11):

| Mean Ejection Velocity (m/s) | Range in *a* (AU) | Range in *e* | Range in *i* (°) |
|---|---|---|---|
| 200 | 40 – 47 | 0.07 – 0.17 | 24 – 32 |
| 300 | 40 – 49 | 0.06 – 0.20 | 24 – 33 |
| 400 | 40 – 51 | 0.05 – 0.22 | 23 – 34 |

We therefore predict that future Haumean family members will be found primarily within this region of the trans-Neptunian realm.

- The orbital diffusion of the stable theoretical family fragments in eccentricity and inclination over billions of years is extremely small (Figs. 4-7). Therefore, the intrinsic orbital distribution of Haumea's family at the present time can be used to draw conclusions about the nature of the collision that originated the family, thought to have occurred some ~4 Gyr ago (e.g., the most likely distribution of fragment ejection velocities).

- If the ejection velocities of the fragments are strongly dependent on their size, the larger fragments will likely be more tightly clustered in orbital element space than their smaller counterparts within the trans-Neptunian belt (Fig. 13). Therefore, future determination of the size distribution of Haumea's family, coupled with the distribution of those fragment in space, will tell us whether the fragments were ejected with velocities following a size dependence, which in turn can provide important information on the physics of collisions between bodies in the outer Solar system.

- The theoretical family fragments were found to populate all four dynamical classes within the trans-Neptunian belt (namely: classical, resonant, scattered and detached TNOs). However, the majority of the surviving fragments fall into the classical and detached classes, whilst only a minority remained within the scattered (<1%) and resonant groups (<10%). The fraction of resonant fragments, in particular, is sensitive to the timing of the collision and the timing and nature of Neptune's orbital evolution during the early Solar system. In this way, future determination of the fraction of Haumea's family members that are trapped in resonances will help in the determination of when and where the family-originating event occurred.

- The orbital distributions of theoretical families obtained after 4 Gyr change very little when massive bodies in the trans-Neptunian region are included in the calculations in addition to the giant planets (Fig. 12). We therefore conclude that the gravitational influence of Haumea, the most massive family members, Pluto and Eris plays no role in determining the spread of the family.

- Approximately 25-40% of the fragments acquired unstable orbits, and were subsequently lost from the Solar system. The main way in which these fragments were removed was through ejection by (mainly) Jupiter, with a secondary sink of material being the collision of fragments with planets. Of these unstable fragments, just ~1.9-2.5% were lost during the last 1 Gyr of dynamical evolution (Fig. 9). Based on these results and optimistic assumptions, we do not expect slowly diffusing unstable fragments to contribute significantly to the currently known populations of Centaurs and short period comets.

- The formation of the Haumea collisional family probably occurred after the bulk of Neptune's migration was complete, and potentially even some time after that migration had ceased completely. However, more work is necessary to confirm this result.

Given our current poor understanding of collision physics for objects in the outer Solar system and the small number of Haumea collisional family members identified thus far, dedicated theoretical and experimental investigations into the collisions of TNOs, and further identification of new Haumea family members will greatly increase our knowledge of the nature of the creation of the Haumean family, and other collisional processes in the Solar system.

In future work, we intend to improve our model of theoretical collisional families and perform more detailed comparisons with observations, including the use of more realistic data of ejection velocities and other key parameters, such as brightness distributions, resonant population characteristics, etc. We also intend to investigate the origin and dynamical evolution of other potential collisional families associated with other dwarf planets and large TNOs, such as Pluto and Eris. Since collisions probably played a crucial role in sculpting the orbital structure of the trans-Neptunian region, future investigations coupling the gravitational perturbations of the planets and massive bodies with the collisional fragmentation of TNOs over billion year timescales will play an important part in the development of a more comprehensive understanding of the origin and evolution of small bodies, satellites and planets, both in our own Solar system and beyond.

## ACKNOWLEDGEMENTS

We would like to thank both an anonymous referee and referee Darin Ragozzine for a number of helpful and detailed comments, which allowed us to improve the overall presentation and flow of this work. PSL gratefully acknowledges the use of the General Purpose cluster at the National Observatory of Japan (NAOJ) to perform the calculations presented in this work. JH acknowledges the financial support of the Australian Research Council, through the ARC discovery grant DP774000. We also thank Michele Bannister for helpful discussions and suggestions.

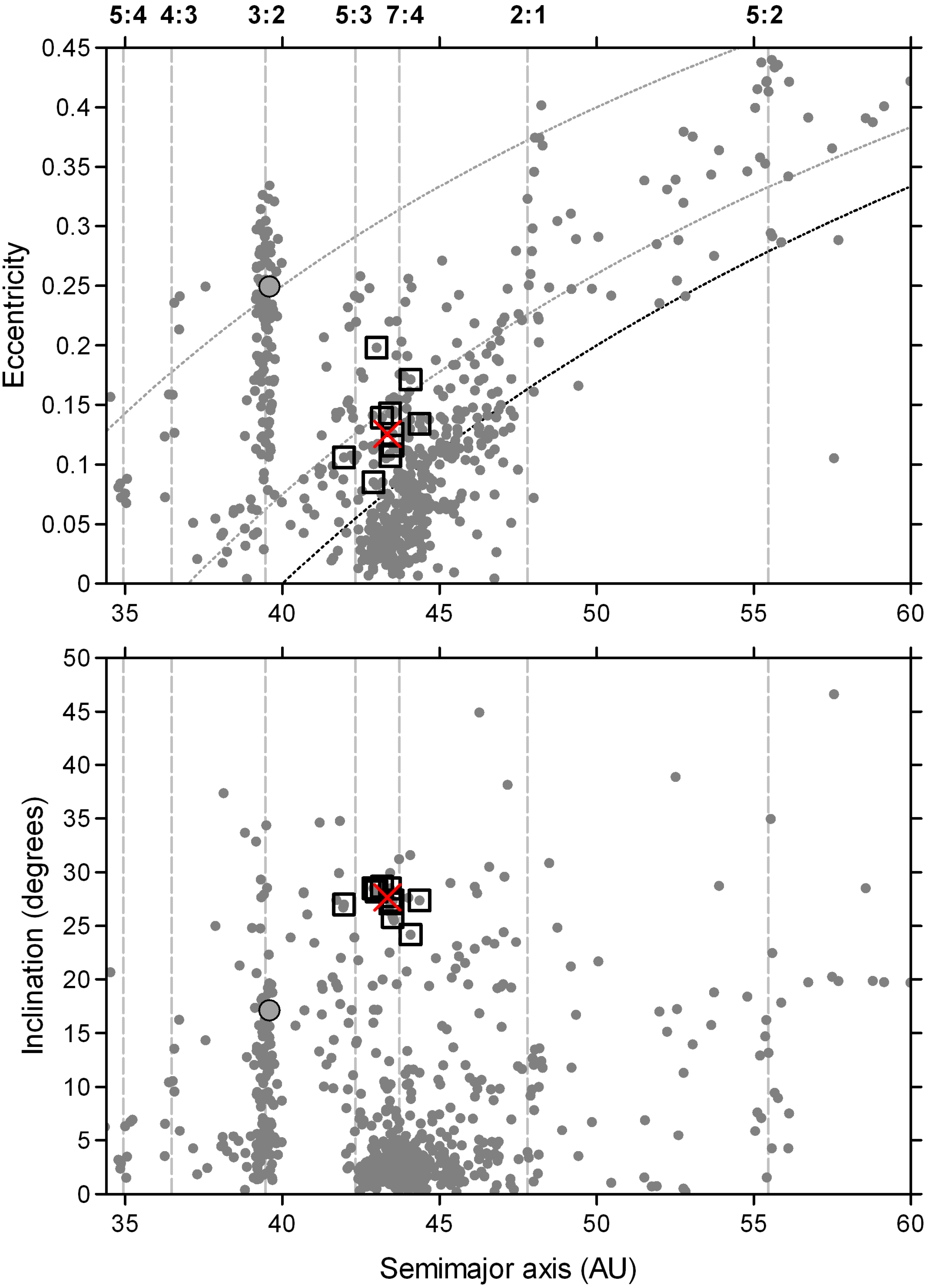

**Figure 1:** The orbits of 759 TNOs (gray circles) taken from the Asteroids Dynamic Site, AstDyS, on 30th September 2010. For clarity, only those objects with orbital uncertainties of ($a_{uncertainty}$ / $a$ ) < 1% (1σ) are shown. Pluto is the gray closed circle. Currently known members of the Haumea collisional family are denoted by squares. The supposed location of the event (a giant impact) that originated the family is marked by the red cross (as determined by Ragozzine & Brown 2007). Perihelion distances of 30, 37 and 40 AU are illustrated by dotted lines (upper panel). The location of Neptunian mean-motion resonances are indicated by vertical dashed lines. The 12:7 resonance is located at ~43.1 AU, between the 5:3 ($a$ ~ 42.3 AU) and 7:4 ($a$ ~ 43.7 AU) resonances, all of which lie within the classical region of the trans-Neptunian belt at ~37-50 AU. The orbits are represented by osculating elements at current epoch.

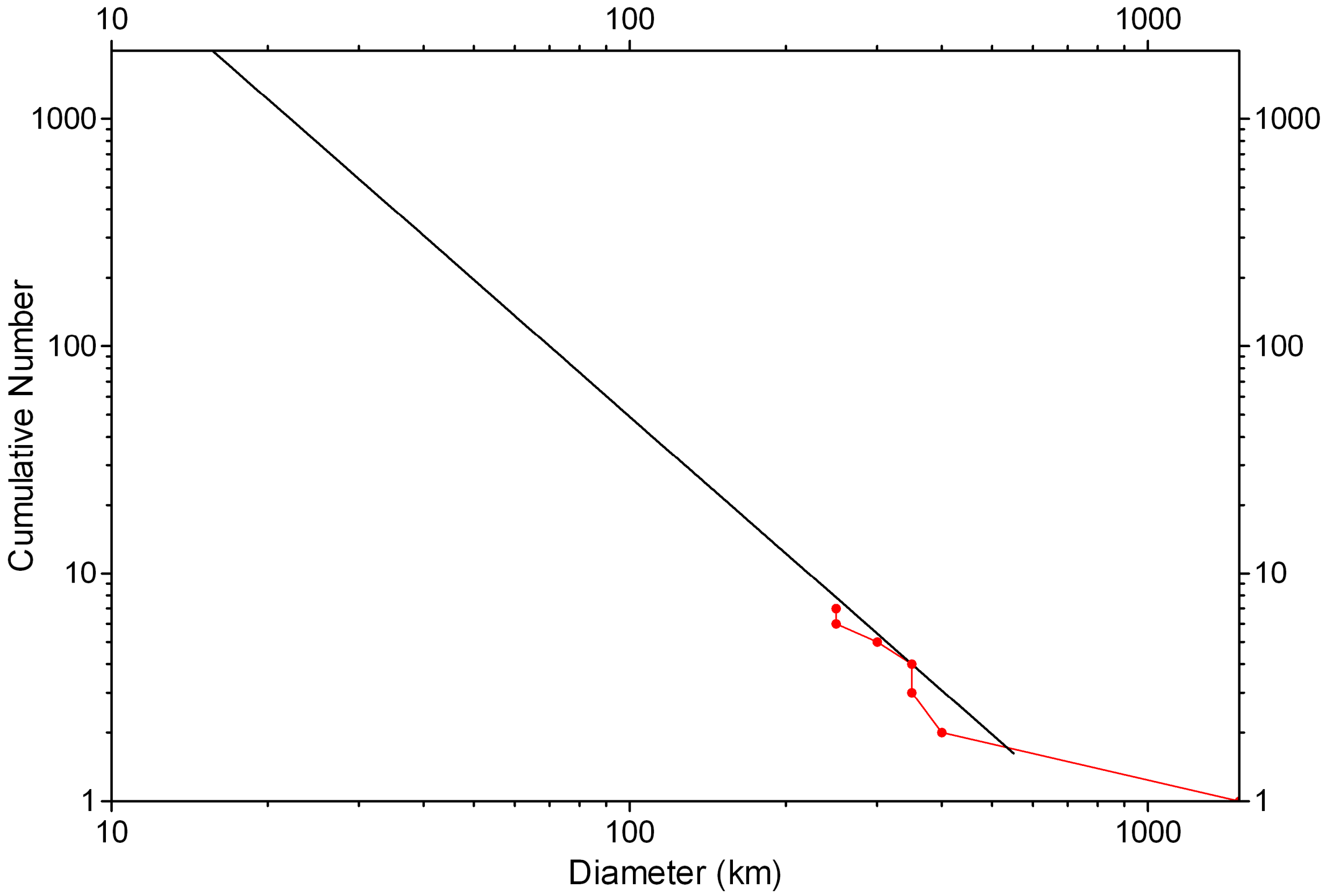

**Figure 2:** The cumulative size distribution assumed for our primordial Haumean collisional family (black line). The distribution is modelled as a power law of slope -2. The six members of the current day Haumean family upon which this tentative distribution is based are shown by red points, connected by a red line. For more details, see Section 2.

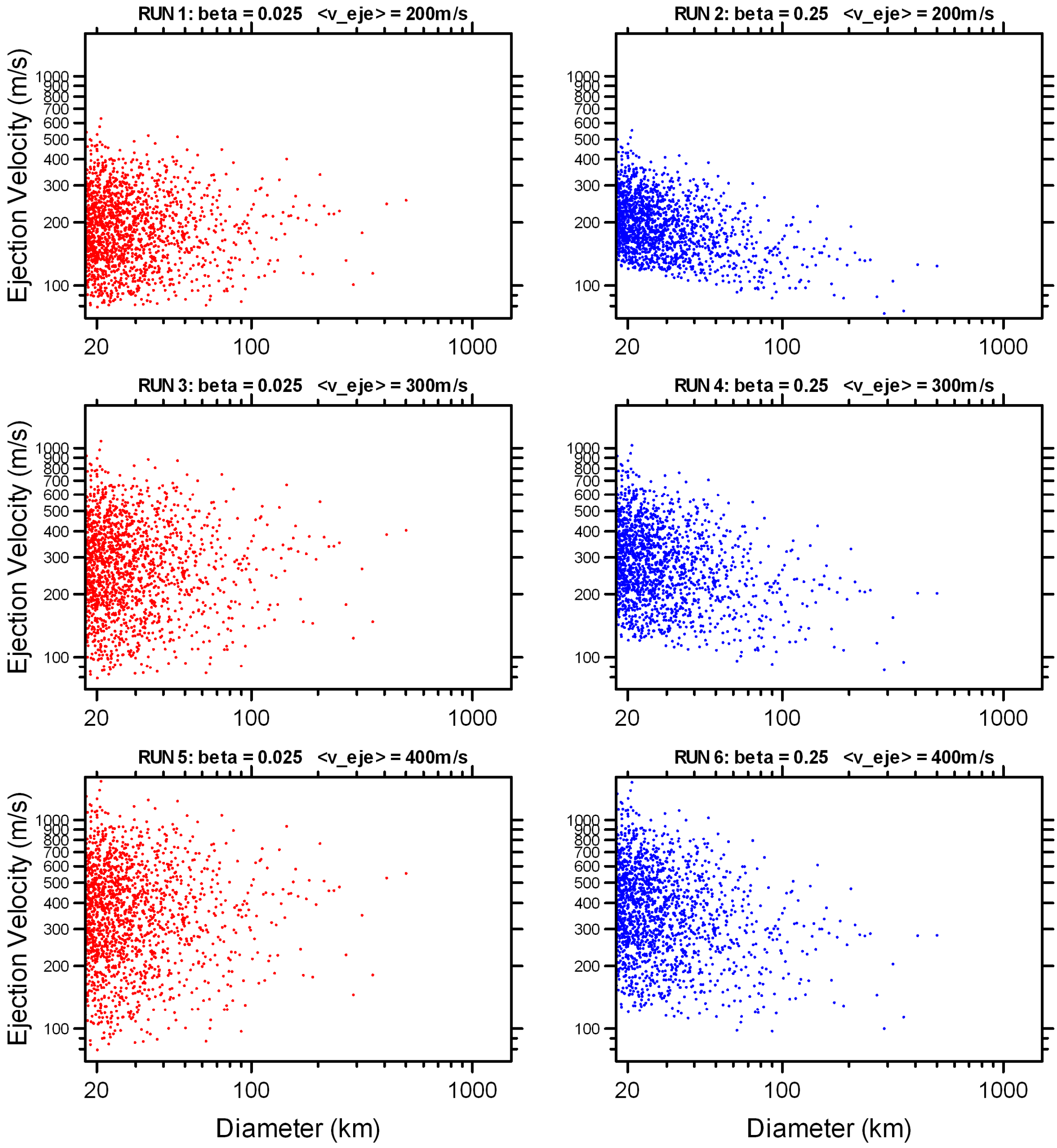

**Figure 3:** The distribution of the ejection velocities of modelled theoretical Haumean collisional family members as a function of their size. These distributions were used in the main runs of the simulations performed in this work (detailed in Table 2).

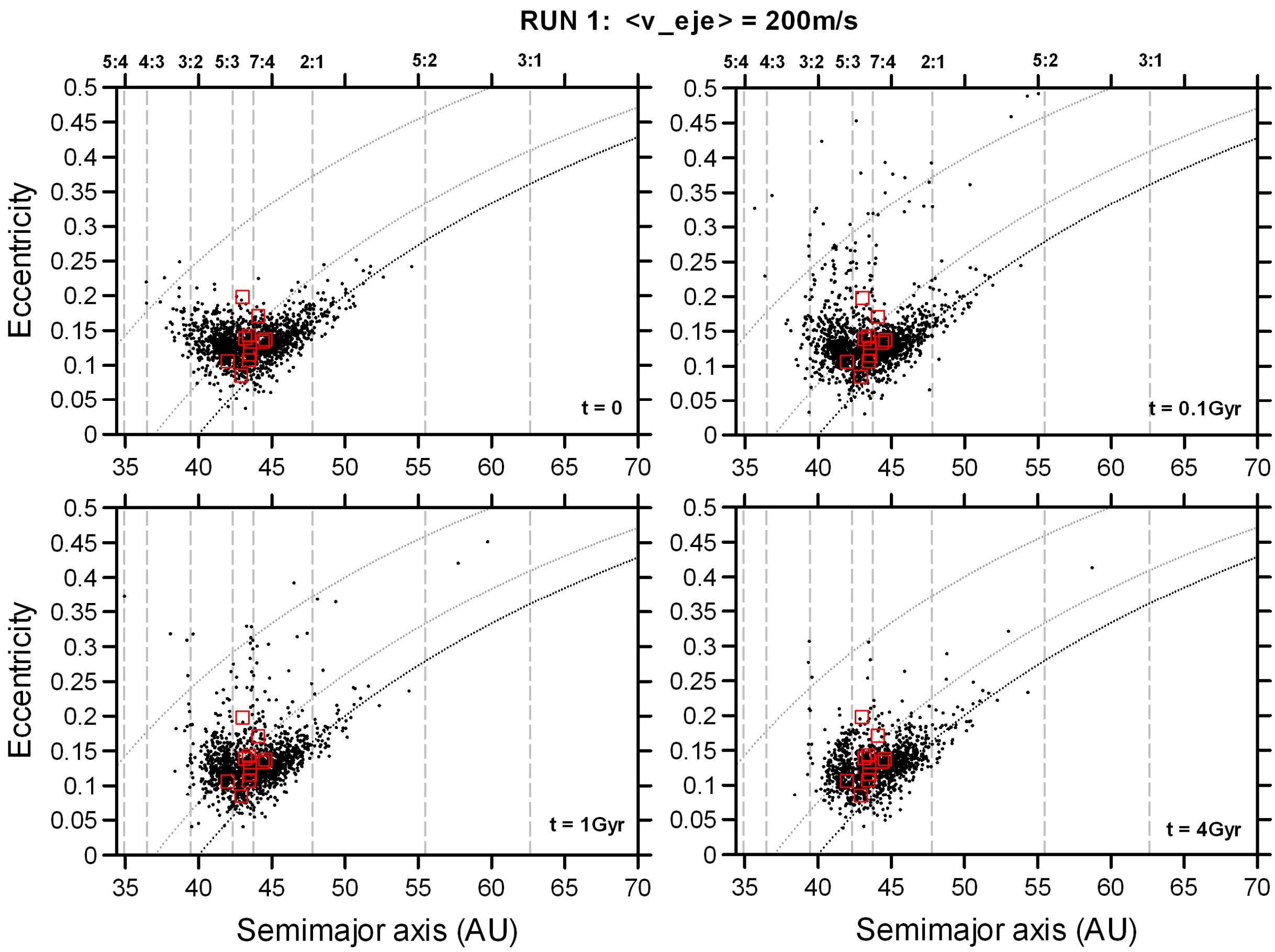

**Figure 4:** Temporal evolution in *a*-*e* element space of a representative theoretical Haumean collisional family modelled with ejecta fragments following a mean ejection velocity of 200 m/s (Run 1; see also Table 2). Currently known members of the Haumean collisional family are shown by red squares. Perihelion distances of 30, 37 and 40 AU are illustrated by dotted lines (left panels), whilst relevant Neptunian mean motion resonances are indicated by their ratios at the top of the figure and vertical dashed lines in all panels. The 12:7 resonance is situated at *a* ~ 43.1 AU, in between the 5:3 (*a* ~ 42.3 AU) and 7:4 (*a* ~ 43.7 AU) resonances. The outcomes of Run 2 and Runs 1a-c were very similar to those shown in this figure. Objects not shown within the limits of the panels follow the "wings" of the overall distribution to smaller and larger semi-major axes, and typically represent less than 1% of the total population (<<1% after 4 Gyr).

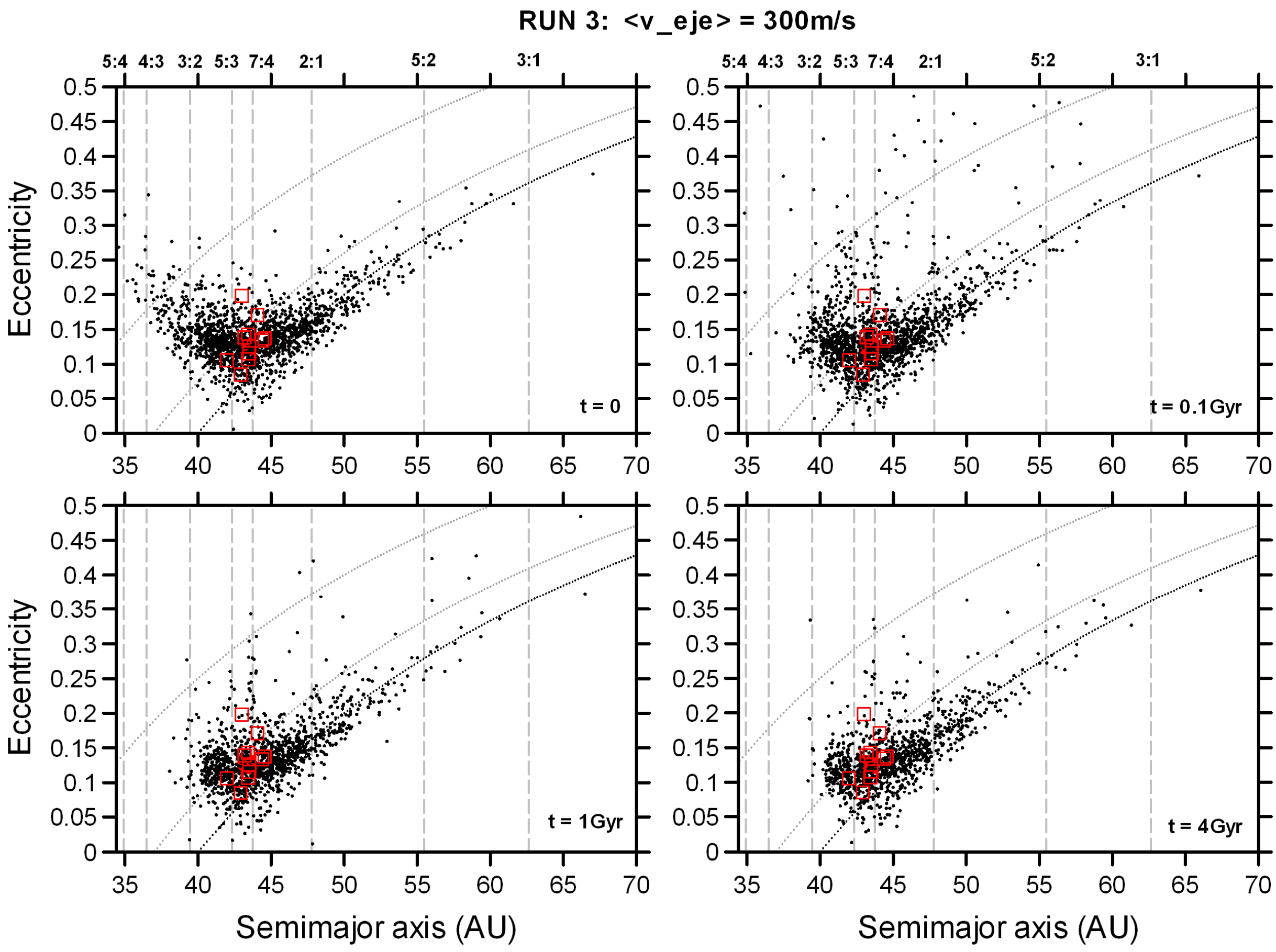

**Figure 5:** Temporal evolution in *a*-*e* element space of a representative theoretical Haumean collisional family modelled with ejecta fragments following a mean ejection velocity of 300 m/s (Run 3; see also Table 2). Currently known members of Haumea's collisional family are shown as red squares. The curves, vertical lines, and remarks about objects not shown within the limits of the figure, are the same as those explained in the caption to Fig. 4. The outcomes of Run 4 and Runs 3a-c were very similar to those shown in this figure.

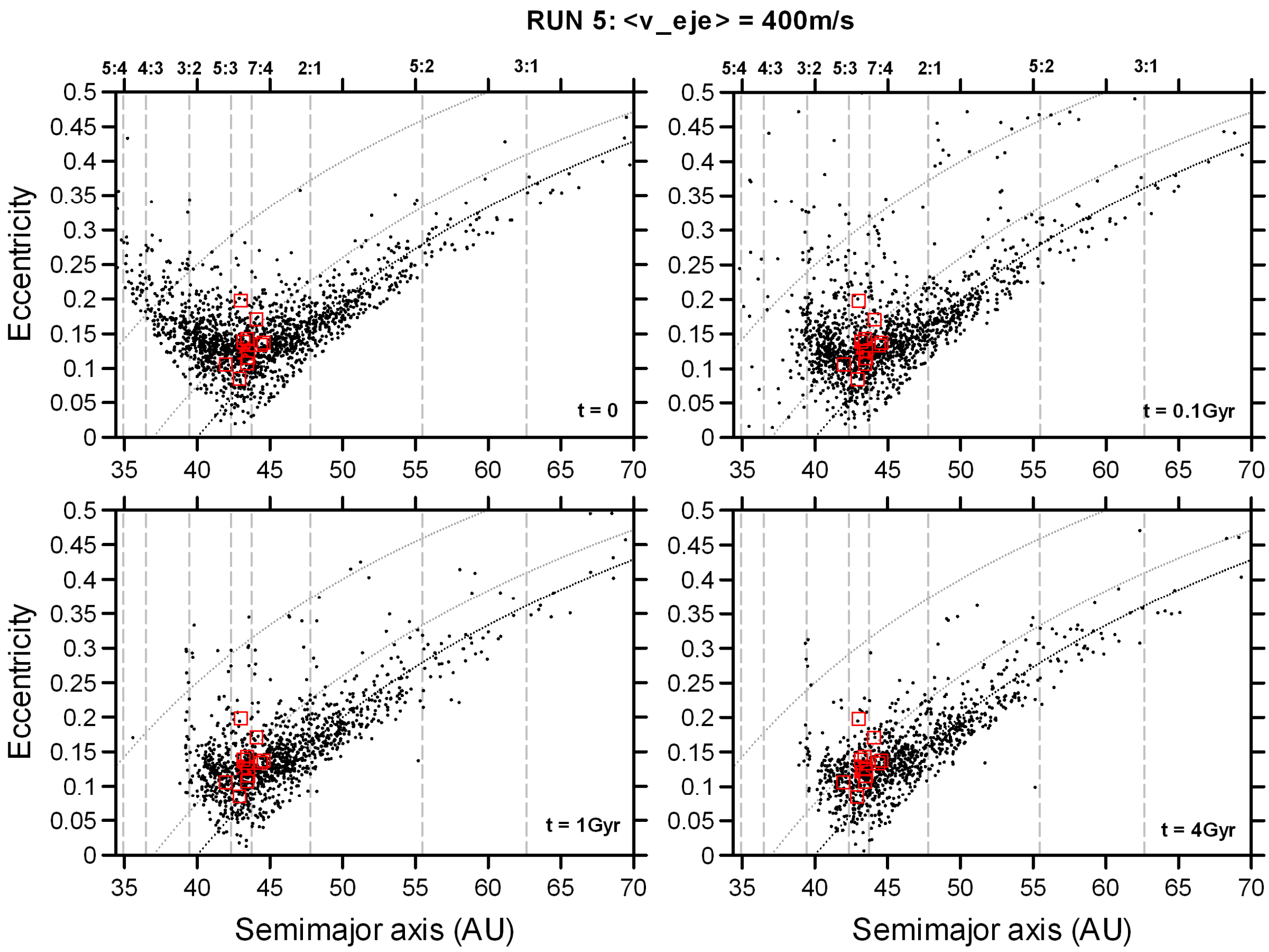

**Figure 6:** Temporal evolution in *a-e* element space of a representative theoretical Haumean collisional family modelled with ejecta fragments following a mean ejection velocity of 400 m/s (Run 5; see also Table 2). Currently known members of Haumea's collisional family are shown as red squares. Objects not shown within the limits of the panels follow the "wings" of the overall distribution to smaller and larger semi-major axes and typically represent less than 5% of the total population (<1% after 4 Gyr). The curves and vertical lines are the same as those detailed in the caption to Fig. 4. The outcomes of Run 6 were very similar to those shown in this figure.

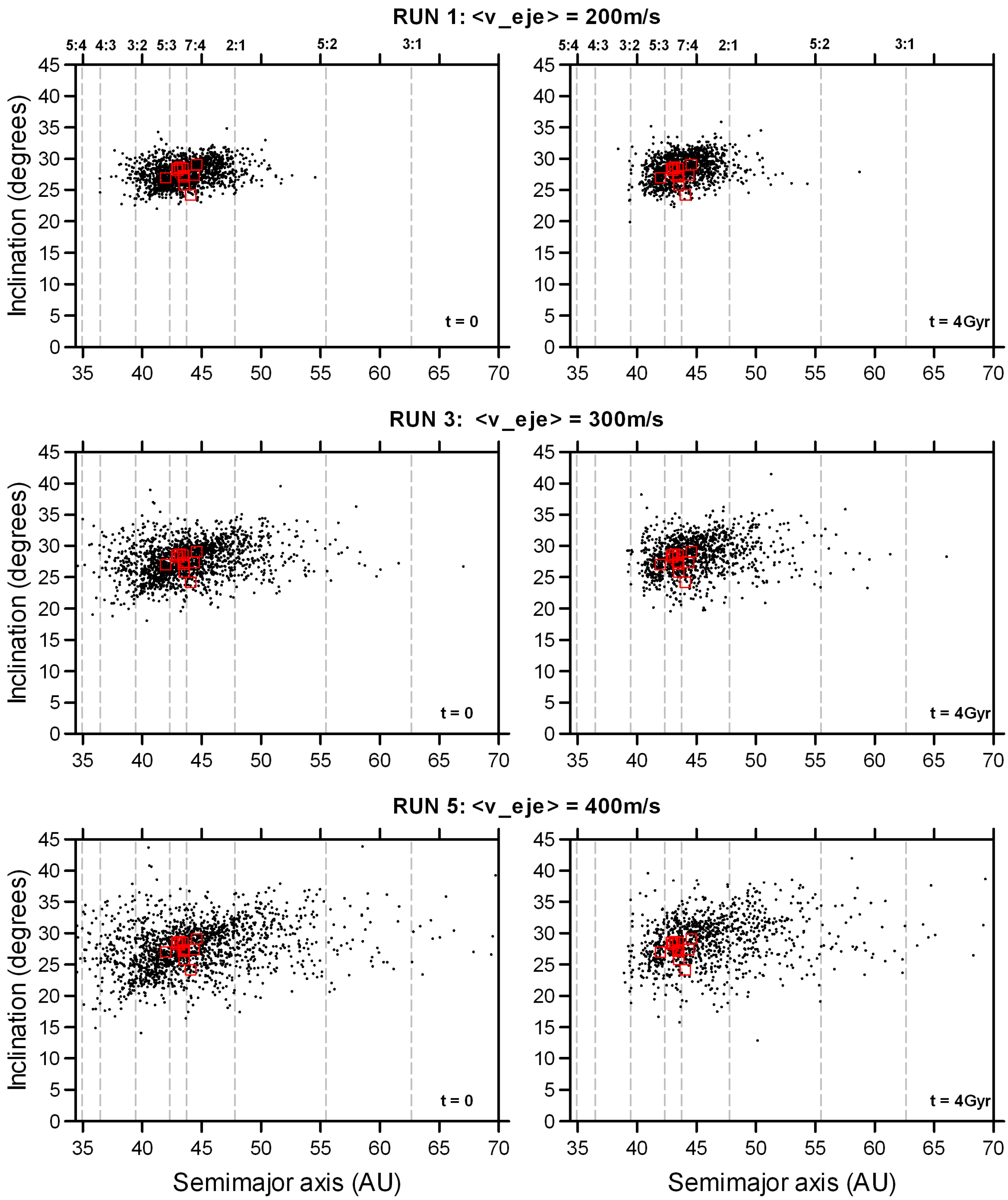

**Figure 7:** Initial conditions (left panels) and the final outcome after 4 Gyr (right panels) of evolution in *a-i* element space of representative theoretical Haumean collisional families modelled with ejecta fragments following a mean ejection velocity of 200, 300 and 400 m/s, respectively (corresponding to our Runs 1, 3 and 5, as described in Table 2). Currently known members of the Haumean collisional family are shown by red squares. Relevant Neptunian mean-motion resonances are indicated by their ratios at the top of the figure and vertical dashed lines in all panels, as explained in the caption to Fig. 4. Objects not shown within the limits of the panels typically represent <1% of the total population after 4 Gyr.

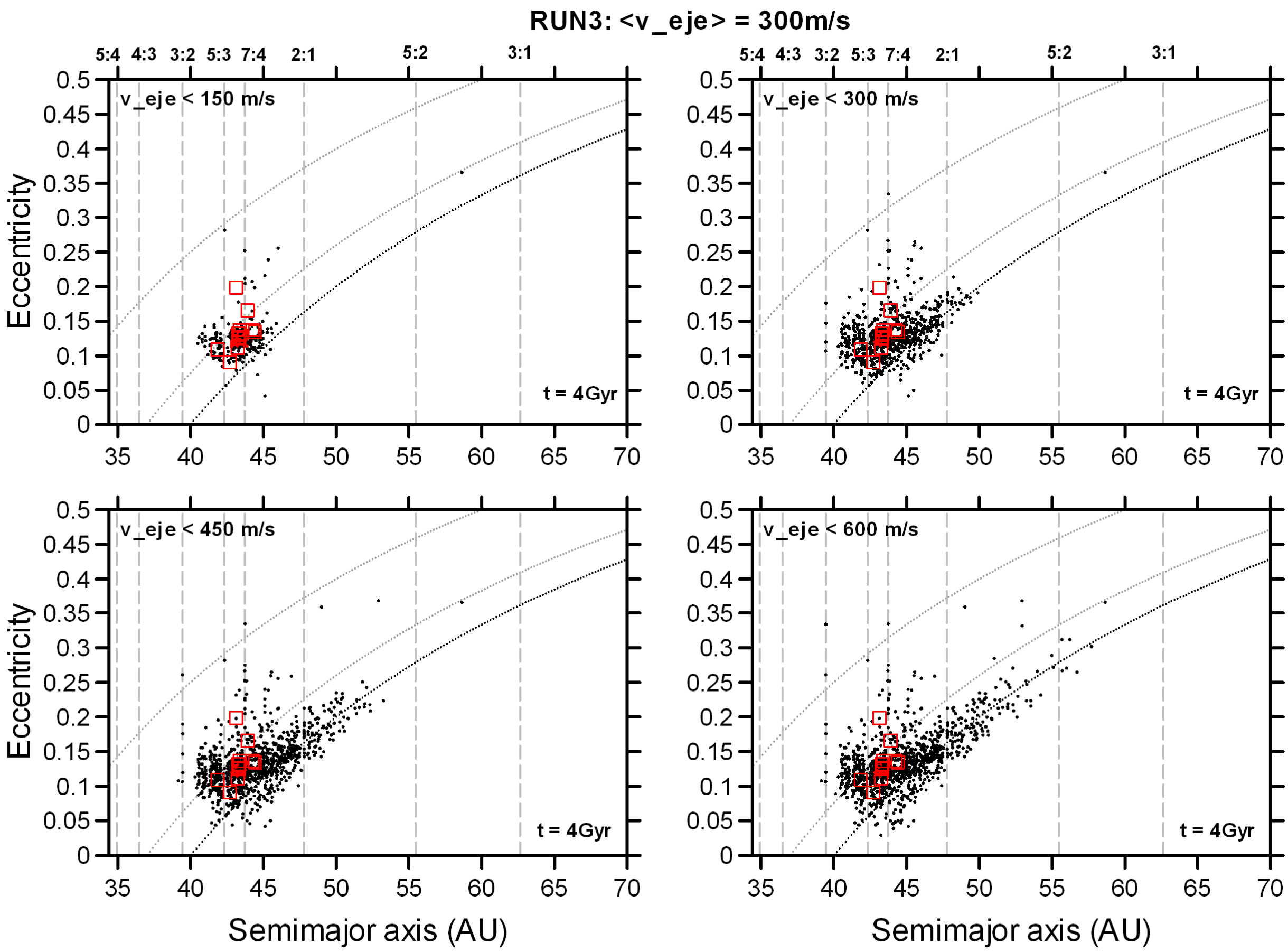

**Figure 8:** The orbits of representative theoretical Haumea collisional family members remaining after 4 Gyr, as a function of initial ejection velocity. The data was taken from Run 3 (see Table 2). The results from other runs are qualitatively similar. Currently known Haumea collisional family members are shown with red squares. The orbits of all objects were averaged over the last 50 Myr of the integrations for more accurate representation of the clustering, and to allow comparison with similar work in the literature. The curves and vertical lines are the same as those described in the caption to Fig. 4.

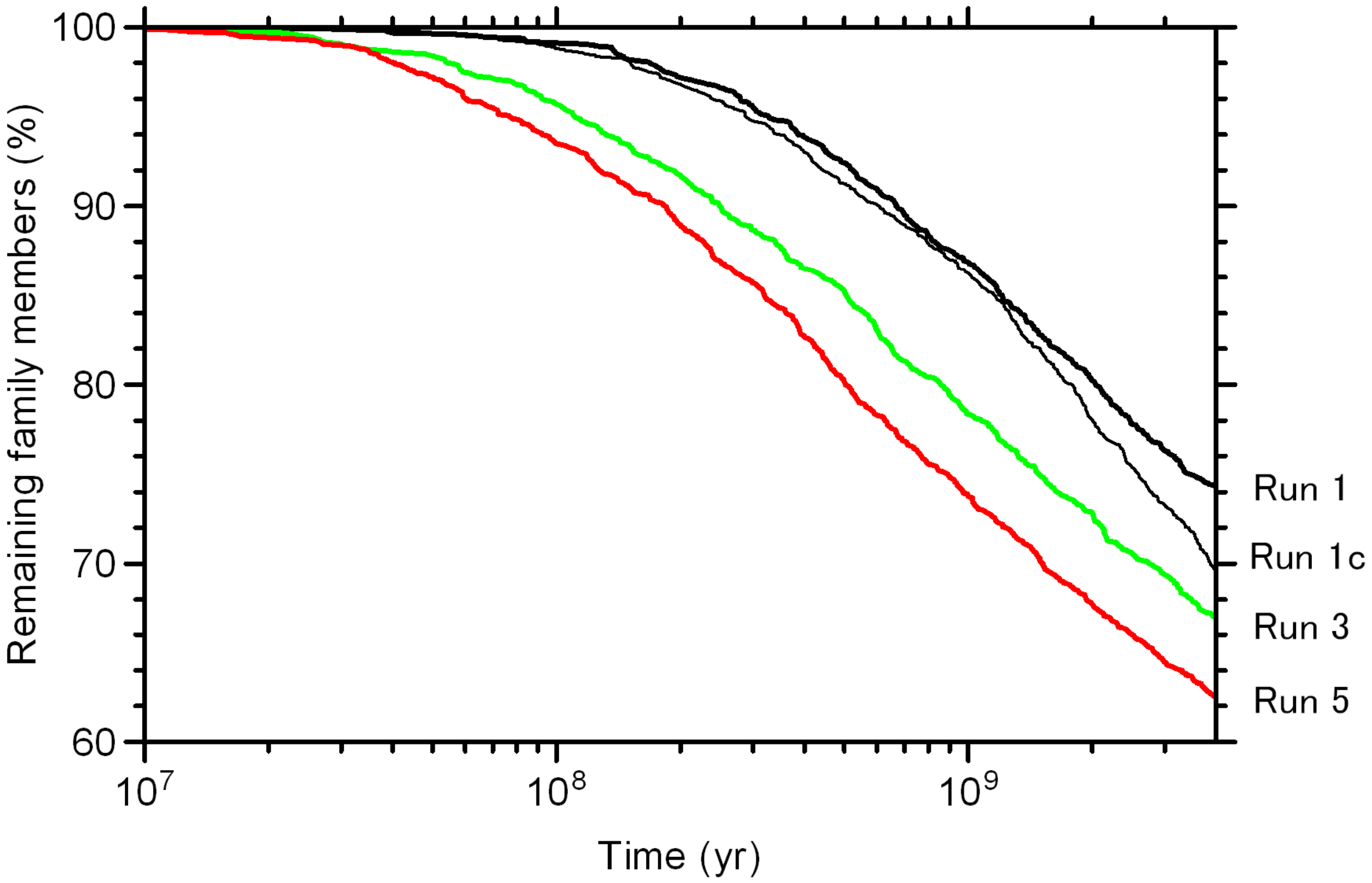

**Figure 9:** The number of Haumean family members remaining in the trans-Neptunian belt as a function of time within our simulations, for scenarios in which the fragments had a mean ejection velocity of 200 m/s (Run 1), 300 m/s (Run 3) and 400 m/s (Run 5). We also added Run 1c, where in addition to the four giant planets, Haumea, the next four largest family members, Pluto and Eris were considered as massive bodies in the simulation. The other scenarios considered in this work yielded qualitatively similar results to that shown here. For more details, see Table 2.

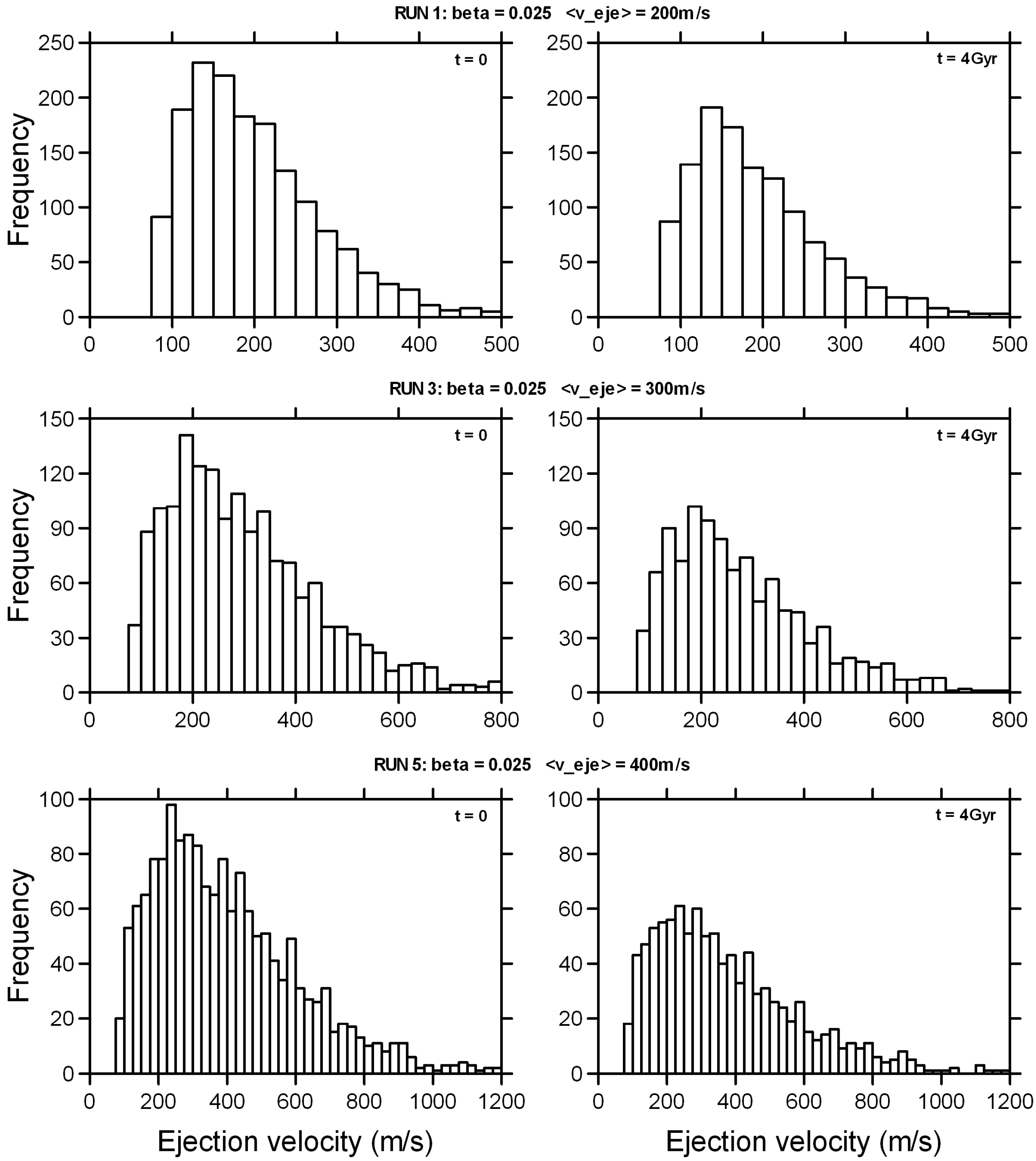

**Figure 10:** Histograms showing the distribution of ejection velocities of fragments for a representative theoretical Haumean collisional family. The distribution of all fragments as modelled in our three main scenarios described by Runs 1, 3 and 5 (see Table 2) at the beginning of the simulations is shown on the left panel, while the distribution of only those particles that remained in the trans-Neptunian belt after 4 Gyr is shown on the right panel. Notice that the ejection velocities shown in the right panels refer to the initial values of the remaining objects, so that they do not represent the ejection velocities that would be observed at the present epoch. The bin size is 25 m/s.

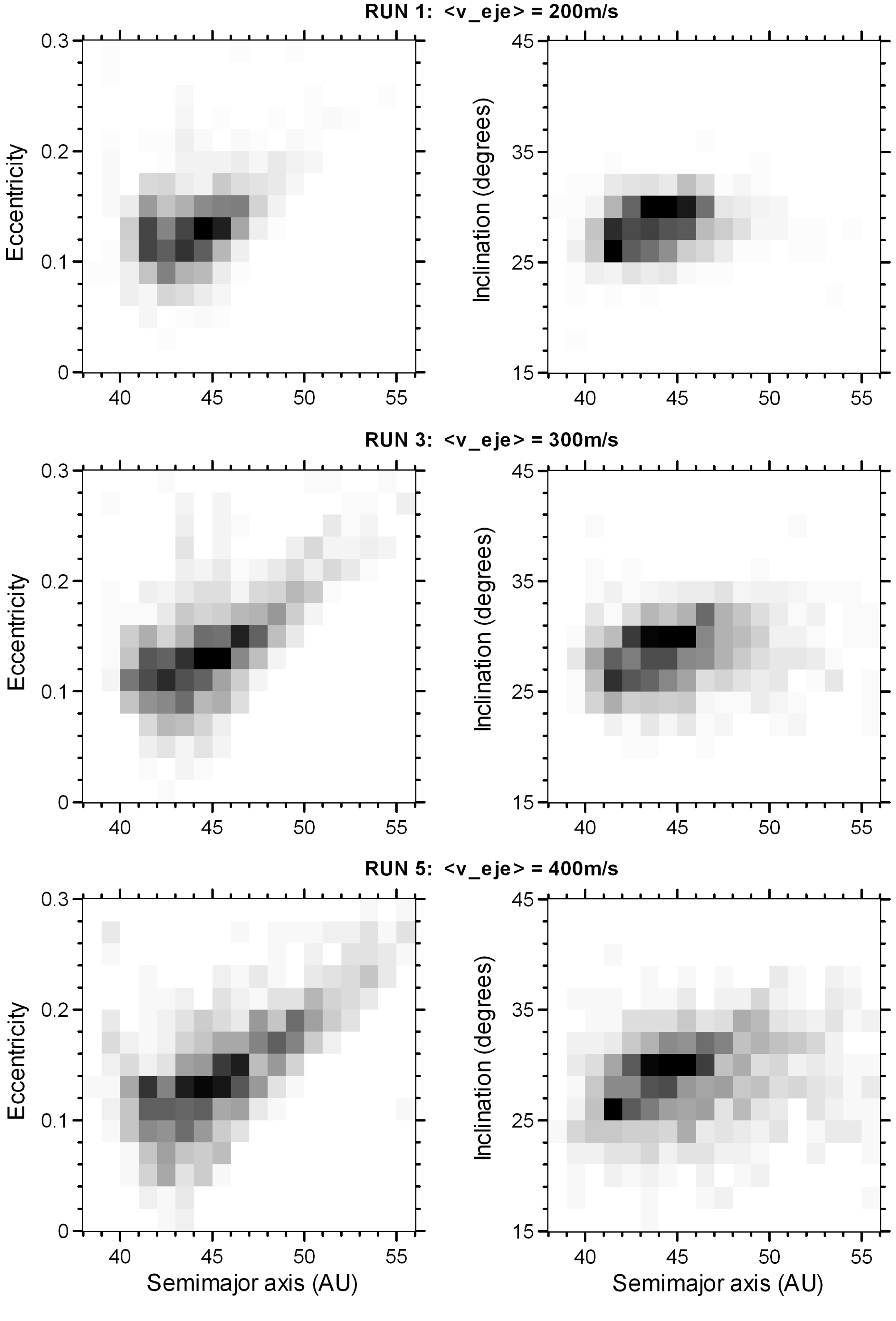

**Figure 11:** Number densities of fragments in *a-e* and *a-i* element space for the three scenarios considered in this work after 4 Gyr of dynamical evolution. The results for collisional families with mean ejection velocities of 200, 300 and 400 m/s are presented (corresponding to our Runs 1, 3 and 5, as described in Table 2). Regions containing different concentrations of fragments are indicated by distinct grey scale shaded regions. The densest region was normalised by the highest number of fragments in a single region for each panel. The darkest and lightest shaded regions typically contain several tens and a few objects, respectively (i.e., roughly an order of magnitude difference). The outcomes for Runs 2, 4 and 6 were very similar to those shown from top to bottom in this figure, whilst the outcomes for Runs 1a-c and 3a-c essentially reproduced those for Runs 1 and 3, respectively (top and middle panels). The orbits of all objects were averaged over the last 50 Myr of the integrations for a more accurate representation of their clustering. Objects not shown within the limits of the panels after 4 Gyr were statistically negligible.

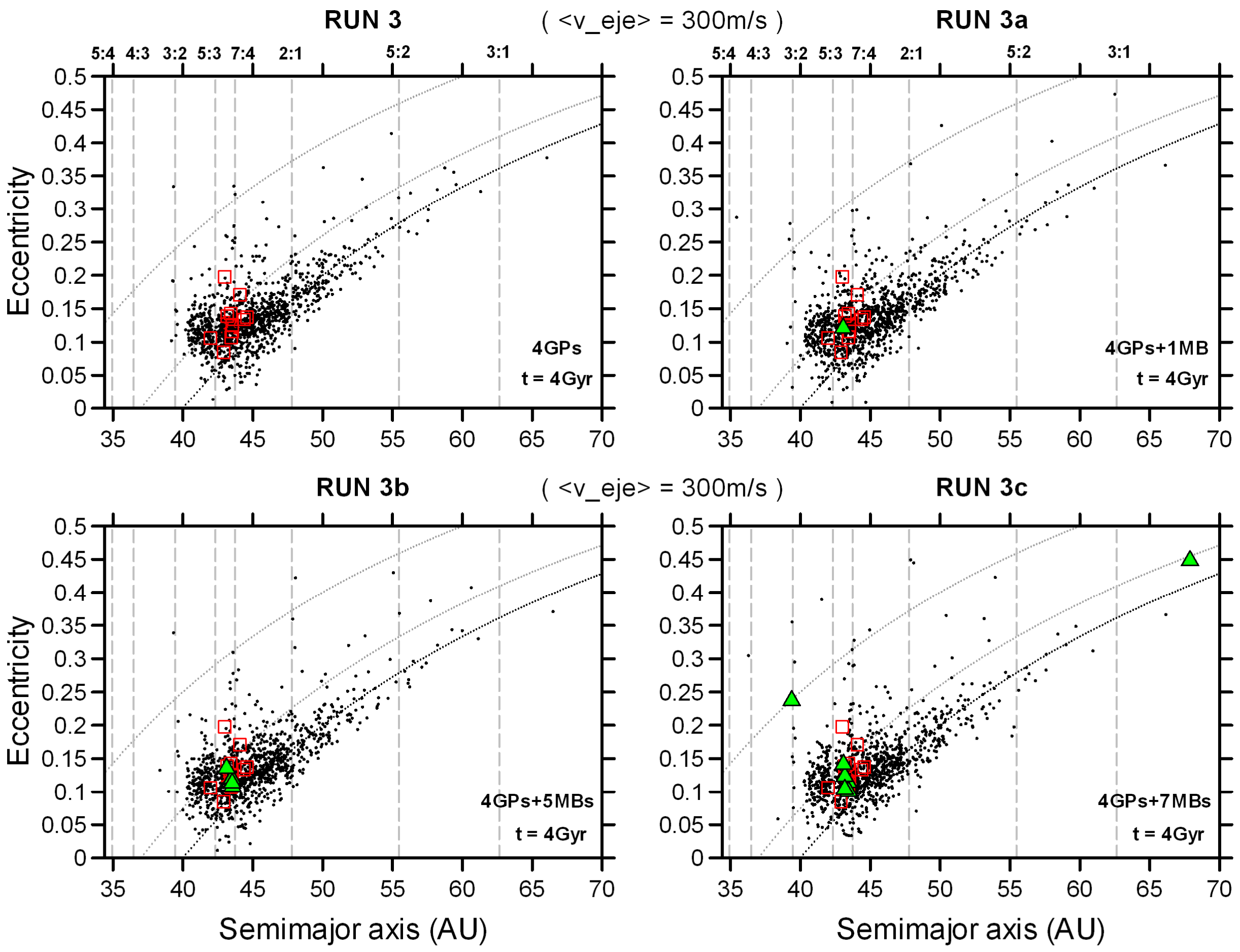

**Figure 12:** The orbital distributions in *a-e* element space of representative theoretical Haumean collisional families after 4 Gyr, modelled with ejecta fragments following a mean ejection velocity of 300 m/s. The simulations followed the evolution of the created family under the gravitational influence of both the giant planets and other massive bodies (MBs) (Runs 3 and 3a-c. See also Table 2). Four cases were considered, each examining a scenario with a different number of MBs: the first scenario considered the influence of just the four giant planets only ('4GPs'), the second utilised the giant planets + Haumea ('4GPs+1MB'), the third considered the giant planets, Haumea, and the next current four most massive family members ('4GPs+5MBs') whilst the final case also incorporated Pluto and Eris, in addition to those objects considered in scenario three ('4GPs+7MBs'). Currently known Haumea collisional family members are shown with red squares. Massive bodies other than the giant planets are represented by green triangles. The curves and vertical lines are the same as those explained in the caption to Fig. 4. Only a small number of objects are not shown within the limits of the panels, and such objects represent a statistically negligible fraction of the total population after 4 Gyr.

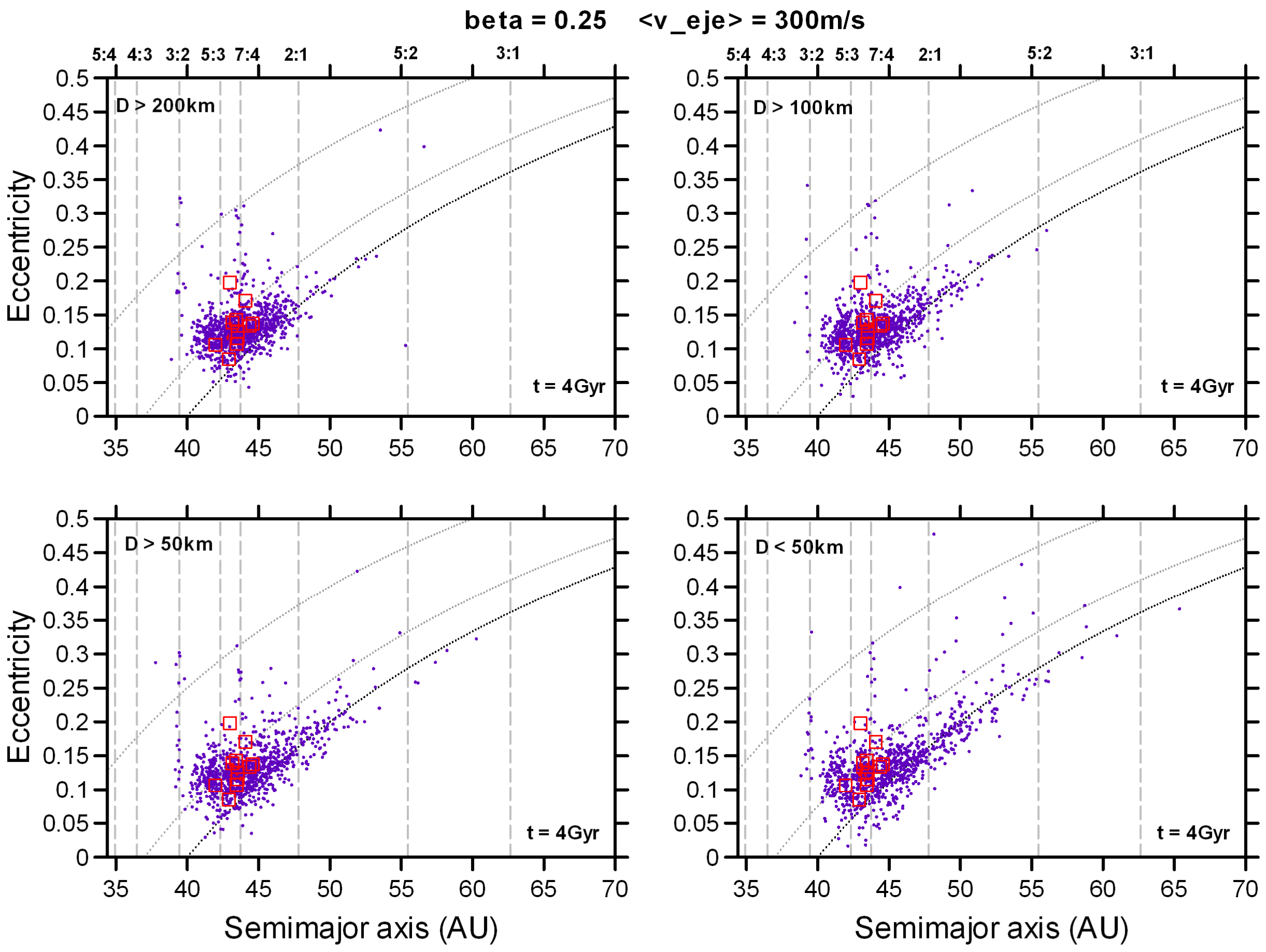

**Figure 13:** The orbital distributions of a representative theoretical Haumea collisional family after 4 Gyr, modelled with fragments whose ejection velocities were inversely proportional to fragment sizes according to Eqs. 2 and 3, with the co-efficient $\beta = 0.25$ (significant dependence of ejection velocity on particle size; Runs 4-200+, 4-100+ and 4-50+. See also Table 2). The initial ejection velocities were used as a proxy for fragment size, so that the latter was not physically incorporated in the orbital integrations. Four cases were considered: fragments larger than 200, 100 and 50 km, and a fourth case with fragments smaller than 50 km. Currently known Haumea collisional family members are shown with red squares. The curves and vertical lines are the same as those explained in the caption to Fig. 4. Only a few objects are not shown within the limits of this figure, and represent a statistically negligible fraction of the total population after 4 Gyr.